\documentclass[aps, twocolumn]{revtex4-1}
\usepackage[utf8]{inputenc}
\usepackage{graphicx}
\usepackage{float}
\usepackage{tabularx}
\usepackage{subcaption}
\usepackage{amsmath}
\usepackage{comment}
\graphicspath{ {/salonisaxena/SKS paper/} }

\begin{document}
\title{Wavenumber selection in pattern forming systems}
\author{S. Saxena}
\affiliation{Department of Physics, Brown University, Providence, Rhode Island 02912, USA}
\author{J. M. Kosterlitz}
\affiliation{Department of Physics, Brown University, Providence, Rhode Island 02912, USA}

\begin{abstract}
Wavenumber selection in pattern forming systems remains a long standing puzzle in physics. Previous studies have shown that external noise is a possible mechanism for wavenumber selection. We conduct an extensive numerical study of the noisy stabilized Kuramoto-Sivashinsky equation. We use a fast spectral method of integration, which enables us to investigate long time behavior for large system sizes that could not be investigated by earlier work. We find that a state with a unique wavenumber has the highest probability of occurring at very long times. We also find that this state is independent of the strength of the noise and initial conditions, thus making a convincing case for the role of noise as a mechanism of state selection. 
\end{abstract}

\maketitle

\section{Introduction} 
This work addresses the question of wavenumber selection in pattern forming systems. Pattern forming systems are characterized by the emergence of a band of spatially periodic steady states, as a certain quantity called the control parameter is varied \cite{cross}. Examples of pattern forming systems in physics are Rayleigh B\'enard convection \cite{cross} and directional solidification \cite{dirsol}. Although a large number of periodicities are mathematically allowed for such systems, experiments and simulations of realistic physical systems have repeatedly shown that only a narrow range of periodicities is realized in practice. The tendency of a system to prefer a narrow set of states out of many possible states is known as wavenumber selection. 

Many mechanisms have been proposed to explain this phenomenon. These include evolution from random initial conditions having a power spectrum centered at a given wavenumber \cite{crgs, schober}, and control parameter ramps \cite{cann, kramer, riecke} (see Section \ref{prev} for a brief discussion). Various computational studies \cite{kers1, kers2, enzo, obeid} have also investigated the role of additive stochastic noise in wavenumber selection. The idea that noise can trigger wavenumber selection can be justified with the help of a simple dynamical system which evolves in a relaxational manner, i.e. it minimizes a potential energy. If the potential has several local minima, the deterministic system will evolve to one of these minima, depending on the initial conditions. However, in the presence of noise, the system will escape from any local minima and eventually reach the global minimum of the potential energy, where it will then spend most of its time, provided the noise strength is small. This has been investigated in \cite{vinals} for the Swift-Hohenberg model \cite{swift}, which is a potential system. While this example of a potential system illustrates wavenumber selection, the systems that are described in \cite{kers1, kers2, enzo, obeid} do \textit{not} have a potential energy function. The role of noise in inducing wavenumber selection in non-potential systems is not well understood. 

In this paper, we further explore the role of noise as a mechanism for wavenumber selection in a model known as the stabilized Kuramoto-Sivashinsky (SKS) equation \cite{misbah}. There are two main reasons for this choice. Firstly, we chose this model because it exhibits rich nonlinear behavior and a band of spatially periodic stationary states, while being relatively simple and one dimensional, which makes simulating it easy. The second reason is that it is non-variational, i.e. the deterministic driving force cannot be expressed as the gradient of a potential. Thus, studying this system would shed light on wavenumber selection in systems where minimization of a potential cannot explain selection of a particular state. 

Noise induced wavenumber selection in this model has been studied before using direct numerical simulation \cite{obeid}, but only for small systems and a limited range of control parameters. More recently, Qiao et al. \cite{qiao} used path integral methods to find the selected wavenumber for a range of control parameter values. It is of considerable interest to extend the work in \cite{obeid} to larger system sizes, since true selection of a unique final state can only occur in the thermodynamic limit. Therefore, our aim in this work is to determine the noise selected wave number for large system sizes and for a wide range of control parameter values through direct integration and to determine if the results of \cite{qiao} can be reproduced in this way. 

This paper is organized as follows: Section \ref{sks} introduces the mathematical formalism used to study pattern forming systems and describes the SKS model. Section \ref{prev} describes in detail the results of \cite{obeid} and \cite{qiao}. Section \ref{simulation} describes our computational method and Section \ref{res} shows some of our results and their interpretation. Finally, Section \ref{conc} discusses some of the drawbacks of our computational methods and touches on potential improvements that will be the focus of future work.

\section{Pattern Formation and the SKS Model} \label{sks}
Pattern forming systems such as the ones mentioned above are represented by nonlinear partial differential equations that govern the evolution of a given physical quantity. A typical equation for such a system is of the form,
\begin{equation} \label{eq:1}
\partial_t u(x,t) =\hat L_pu(x,t) +\hat N[u(x,t)]
\end{equation}
where $u(x, t)$ is a field representing the quantity of interest, $\hat L_p$ is a linear differential operator acting on $u(x,t)$ and $\hat N[u(x,t)]$ is a nonlinear operator. The subscript $p$ on the linear operator indicates that it depends on the control parameter $p$. As an example, in the case of directional solidification, the quantity of interest $u(x,t)$ is the position of the interface between the  liquid and solid phases. The trivial solution (or base state) $u_b(x, t)=0$ is generally a stationary state of these equations. However, this solution is stable only for a certain range of values of $p$. After $p$ crosses a critical value $p_c$, the trivial, spatially uniform state becomes unstable to periodic perturbations and a band of stable, periodic steady states emerges. To determine when the uniform state becomes unstable, we imagine adding to the uniform base state $u_b(x,t)=0$ the perturbation $\delta u \sim e^{iqx+\sigma (q) t}$. This perturbation is periodic in space with wavenumber $q$ and grows with time at a rate $\sigma$. We then substitute $u(x,t) = u_b(x,t) + \delta u$ in Eq. (\ref{eq:1}), retain only terms linear in $\delta u$ and derive an expression for the growth rate $\sigma$ as a function of $q$. If, for a given value of $p$, $Re(\sigma(q))$ is negative, that means that the perturbations with those wavenumbers decay exponentially with time and hence the uniform base state $u_b$ is stable to those perturbations. However, if $Re(\sigma(q))$ is positive, then perturbations with wavenumber $q$ grow exponentially with time, implying that the base state is unstable to them. Of course, in practice, the unbounded exponential growth of these perturbations is balanced by the nonlinear terms in the original equation.

To make these ideas more concrete, we illustrate the above steps for the deterministic SKS model. The SKS equation is given by,
\begin{equation} \label{eq: 2}
\partial_t u(x,t) = -\alpha u(x,t) - \partial_x^2 u(x,t) -\partial_x^4 u(x,t) + {(\partial_x u(x,t))}^2
\end{equation}
Here, $\alpha$ plays the role of the control parameter, and $u(x, t)$ is a dimensionless field of dimensionless space-time variables. This equation is used to describe directional solidification \cite{misbah} and the Burton-Cabrera-Frank model of terrace growth \cite{misbah2}. The trivial solution $u_b(x,t) = 0$ is one of the steady states of this equation. We now add to this solution a small perturbation of the form $\delta u \sim e^{iqx+\sigma (q) t}$. (We restrict ourselves to the case where $\sigma$ is purely real.) Substituting into Eq. (\ref{eq: 2}) and linearizing about the base state $u_b(x,t)=0$ state gives us an expression for the growth rate,
\begin{equation} \label{eq:3}
\sigma = -\alpha + q^2 - q^4
\end{equation}

\begin{figure}
\centering
\fbox{\includegraphics[width=0.48\textwidth]{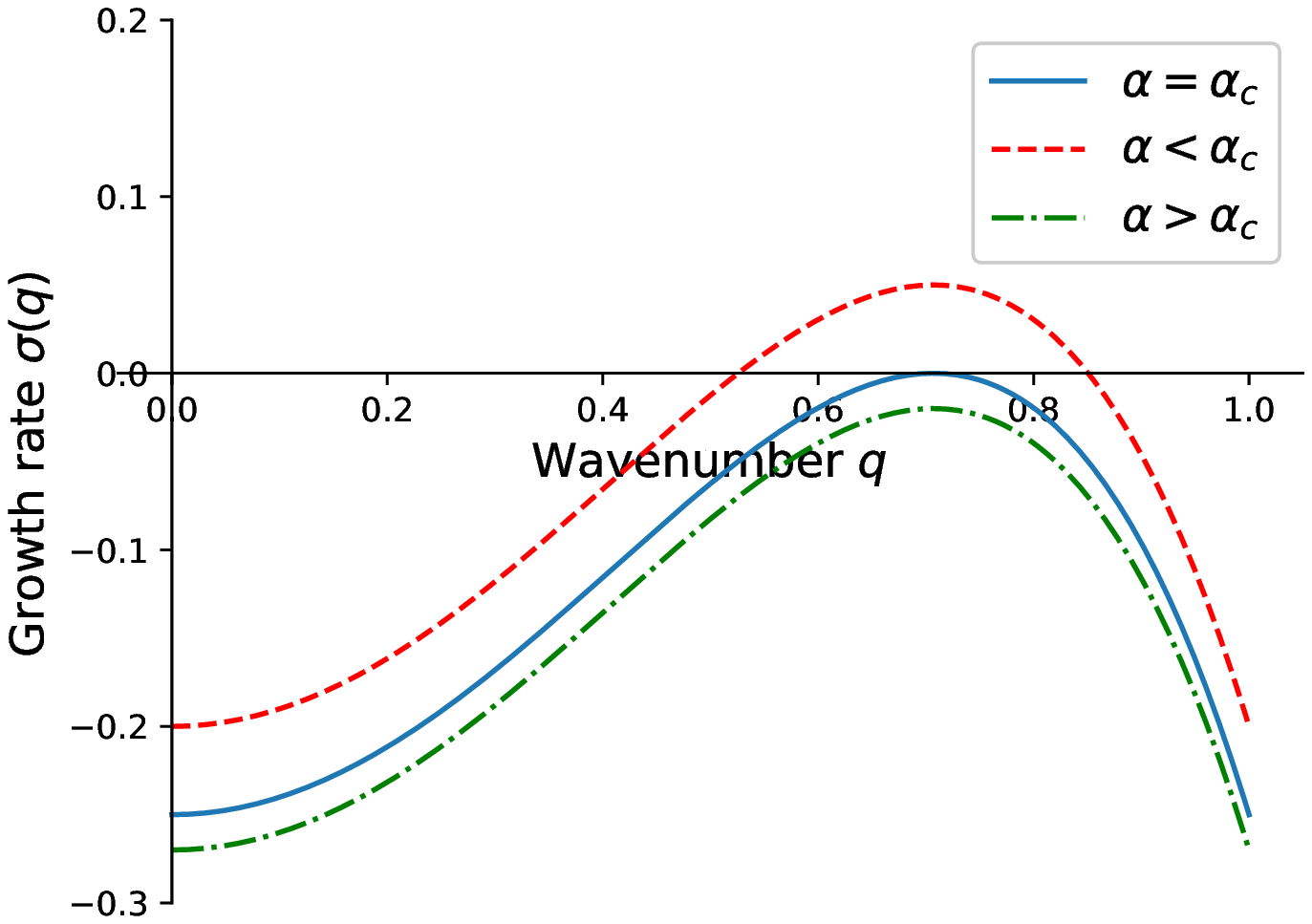}}
\caption{Growth rate of periodic modes for values of $\alpha$ below, above and at the threshold value $\alpha_c=0.25$}
\label{fig: 1}
\end{figure}
From this, we see that the growth rate is non-negative for $1/2-\sqrt{1/4-\alpha} \leq q^2 \leq 1/2+\sqrt{1/4-\alpha}$ and $\alpha \leq 1/4$, as shown in Fig. \ref{fig: 1}. 
 For these modes, the state $u_b(x,t)=0$ is unstable. By solving the equation $\frac{\partial \sigma}{\partial q}=0$, we see that the growth rate is maximum for $q=q_c=1/\sqrt{2}$, for all $\alpha$. In summary, for $\alpha \leq 1/4=\alpha_c$, a band of periodic steady states exists, with wavenumbers given by,
\begin{equation} \label{eq: 4}
1/2-\sqrt{1/4-\alpha} \leq q^2 \leq 1/2+\sqrt{1/4-\alpha}, \ \ \ \ \  \alpha \leq \alpha_c
\end{equation}
These periodic states may themselves be unstable to long wavelength periodic perturbations. Hence, in practice, one observes a band of periodic states that is narrower than suggested by Eq. (\ref{eq: 4}) and is called the Eckhaus stable band \cite{crgs}. Wavenumbers within this band are stable to long wavelength periodic perturbations.

\section{Previous Studies of Wavenumber Selection} \label{prev}
As mentioned in the introduction, many possible explanations for wavenumber selection have been suggested. Schober et al. \cite{schober} studied wavenumber selection in the context of the one dimensional Swift-Hohenberg equation \cite{swift}. The deterministic equation was integrated numerically, starting from random initial conditions with a power spectrum centered at a particular wavenumber $\bar{k}$. It was observed that the width of the noise averaged power spectrum (or the structure function) decreased with time until a sharp power spectrum centered about a final wavenumber $k_{\infty}$ was obtained. Their simulations showed that the final wavenumber depended on the value of $\bar{k}$.

Another proposed mechanism is the slow spatial variation of the control parameter, so that it changes from a value below threshold in one part of the experimental apparatus to a value above threshold in another part of the apparatus. This has been investigated experimentally \cite{cann} for the case of rotating Couette-Taylor flow. The experimental apparatus consists of two coaxial cylinders, separated by a gap that contains fluid. The inner cylinder is allowed to rotate with a given angular velocity. The control parameter, which is a function of the gap, is varied gradually by reducing the gap linearly towards the bottom of the apparatus. The variation in the gap is characterized by the angle made by outer wall with the vertical. The wavelength of the resulting flow pattern was measured in the straight section for different bulk values of the control parameter and was observed to lie in a narrow band, whose width decreased as the control parameter was increased beyond threshold. The observed wavelength was also found to be a periodic function of the aspect ratio (length of straight section/gap). The role of ramps has also been studied analytically \cite{kramer, riecke}. Ramps provide an efficient way to obtain a desired, well defined periodic state. However, it must be noted that ref. \cite{riecke} shows that varying different quantities which influence the control parameter leads to different selected wavenumbers. This implies that the selected wavenumber depends either on the initial conditions, or in case of control parameter ramps, on the chosen ramping protocol. In other words, the selected wavenumber in a deterministic system is not an intrinsic property of the system itself, but depends on the measuring protocol.

The role of noise in inducing wavenumber selection is the subject of some debate. While thermal noise is small enough that its effect on macroscopic patterns can be neglected \cite{cross}, the same cannot be said about noise in experimental apparatus \cite{kers1, kers2, enzo, obeid, ghs93, pvbr96}. As argued in \cite{kers2}, it is hard to precisely control experimental conditions in situations such as directional solidification, leading to uncertainties which are most easily modeled as additive noise. Studies have also shown that \textit{multiplicative} noise can have non-trivial consequences on the dynamics of bifurcations, such as shifting the primary bifurcation point in the Swift-Hohenberg equation \cite{ghs93} and inducing new patterns which are absent in the deterministic system \cite{pvbr96}. In addition, \cite{kers1, kers2, enzo, obeid} have shown evidence of additive noise induced wavenumber selection. In our view, the other mechanisms mentioned above are too deterministic because they only sample a restricted set of perturbations. In contrast, stochastic noise is the most unbiased mechanism possible because it perturbs all periodic states of the deterministic system equally and does not favor a particular wavenumber over the others. It also samples a much larger set of perturbations. Hence, it is our view that more work needs to be done to gain a deeper understanding of the effects of noise, whether additive or multiplicative.

For the particular case of the SKS equation, Obeid et al. \cite{obeid} and Qiao et al. \cite{qiao} have recently investigated noise induced wavenumber selection. We present here a brief summary of their results, which are relevant to this work. Obeid et al. carried out direct numerical simulations of the noisy SKS equation, 
\begin{equation} \label{eq: 5}
\partial_t u(x,t) = -\alpha u - \partial_x^2 u(x,t) -\partial_x^4 u(x,t) + {(\partial_x u(x,t))}^2 + \zeta(x,t)
\end{equation}
where $\zeta(x,t)$ is an additive Gaussian noise satisfying,
\begin{equation} \label{eq: 6}
\langle \zeta(x,t) \rangle = 0  
\end{equation}
and
\begin{equation} \label{eq: 7}
\langle \zeta(x,t) \zeta(x^\prime, t^\prime) \rangle = 2 \varepsilon \delta (x-x^\prime) \delta (t-t^\prime)
\end{equation}
Here $\varepsilon$ is the noise strength. Eq. (\ref{eq: 5}) was discretized using a simple finite difference scheme and the time integration was performed using the explicit forward Euler method.
Various values of $\alpha$ between 0.17 and 0.24 and noise strengths ranging from $10^{-5}$ to $10^{-3}$ were studied. For $\alpha=0.24$, it was found that the state with $q=0.6995$ was most stable, in the sense that the system could not be knocked out of this state for $10^8$ time steps, with noise strengths up to $\varepsilon=5 \times 10^{-4}$. 

For $\alpha$ values farther from the critical value of 0.25, no unique state could be identified as being most stable. For $\alpha = 0.2$, the allowed wavenumbers are $0.589 \leq q \leq 0.767$. It was found that states with $0.650 \leq q \leq 0.712$ all remained stable up to $10^8$ time steps and noise strengths up to $\varepsilon = 2 \times 10^{-3}$. Similarly, for $\alpha = 0.17$, states with $0.650 \leq q \leq 0.699$ were found to remain stable at $10^8$ time steps and noise strengths up to $\varepsilon = 2 \times 10^{-3}$. Increasing the noise strength made it hard to discern any periodic states and it was concluded that very long computational times would be required to destroy the stability of the states. In summary, the simulations in \cite{obeid} hinted that the noise does indeed select a narrrow band of wavenumbers over others. However, selection of a unique state could only be demonstrated for $\alpha=0.24$.  A conclusive numerical demonstration of state selection therefore requires integration over large system sizes, long times and a wider range of control parameters.

A more recent study of state selection in the SKS equation was conducted by Qiao et al. \cite{qiao}. They used the least action principle of Freidlin-Wentzell theory \cite{freidlin} to calculate transition probabilities between pairs of periodic steady states for the SKS equation.  For the following stochastic process defined on a spatial domain [0, L], 
\begin{equation} \label{eq: 8}
\dot \Phi(x,t) = f[\Phi(x,t)] + \zeta(x,t)
\end{equation}
the probability of a particular trajectory $\Phi(x,t) = \phi(x,t)$ defined over a time interval [0, T] is,
\begin{equation} \label{eq: 9}
P_T[\phi(x,t)] = \mathcal{N} \ \text{exp}[-S_T[\phi(x,t)]/\varepsilon]
\end{equation}
where $\mathcal{N}$ is a normalization constant independent of $\phi(x,t)$, $\varepsilon$ is the noise strength defined by,
\begin{equation} \label{eq:10}
\langle \zeta(x,t) \zeta(x^\prime, t^\prime) \rangle = 2 \varepsilon \delta (x-x^\prime) \delta (t-t^\prime) 
\end{equation}
and $S_T[\phi(x,t)]$ is the action, given by,
\begin{equation} \label{eq: 11}
S_T[\phi(x,t)] = \frac{1}{2} \int_0^T \text{d}t \int_0^L \ \text{d}x { [\dot \phi(x,t) -f[\phi(x,t)]]}^2 
\end{equation}
Eq. (\ref{eq: 9}) implies that the most probable trajectory connecting two states of the system is the one which minimizes the action. The most likely paths entering and leaving successive periodic states of the SKS equation were computed by finding the minimum action for transitions between those states (for example, $S^*_{q_j \rightarrow q_{j+1}}$ is the minimum action to go from a periodic state with wavenumber $q_j$ to one with wavenumber $q_{j+1}$, and $S^*_{q_{j+1} \rightarrow q_j}$ is the minimum action for the reverse transition). These values were then used to determine the net direction of transitions between two adjacent states. By implementing this procedure for all pairs of successive steady states, the wavenumber corresponding to the selected state was found. In order to minimize the actions  $S_{q_j \rightarrow q_{j+1}}$ and $S_{q_{j+1} \rightarrow q_j}$, the saddle state between the states $q_j$ and $q_{j+1}$ was found numerically. The saddle state of the lowest order amplitude equation \cite{cross, zimm} was used as the initial guess for this procedure. Having found the saddle state, the time reversed deterministic paths between the saddle state and the states $q_j$ and  $q_{j+1}$ were used as initial guesses to find the minimum action paths (see sec. III B of \cite{qiao} for a detailed discussion).
\begin{table*}[t!]
\caption{\label{table: Table 1} Most visited wavenumbers for various sizes, $\alpha=0.22$.}
\begin{ruledtabular} 
\begin{tabular}{lll}
 
 Number of lattice points $N$ & System Length $(L=Nh)$ & Most visited state \\ 
 \hline
 1024 & 512 & 0.6749 $(n=55)$ \\
 
 1600 & 800 & 0.6754 $(n = 86)$ \\ 
 
 2200 & 1100 & 0.6740 $(n = 118)$\\ 
 
 3000 & 1500 & 0.6744 $(n = 161)$\\
 
 4000 & 2000 & 0.6754 $(n = 215)$\\

 \end{tabular}
 \end{ruledtabular}
 \end{table*}

\section{Calculating the Empirical Probability Distribution of Final States} \label{simulation}
Although Qiao et al. have devised a way to calculate the selected wavenumber, we wish to know if the same results can be obtained by direct integration of the equation of motion. There are two hurdles that must be overcome in order to do so. The first is that farther from threshold, several neighboring states in the middle of the Eckhaus band become very stable to noise \cite{obeid}. In order to determine which one, if any, is the most stable, one has to induce transitions between these states, so that the state in which the system spends most of its time is the most stable state. Observing such transitions between highly stable states requires extremely long integration times, as noted in \cite{obeid}. The second hurdle is that true state selection would only occur in the thermodynamic limit, which necessitates simulation of large systems. These two hurdles suggest that one looks for a fast and efficient integration algorithm. Here, we use a semi-implicit, Fourier spectral integration method \cite{chen}. Using a semi-implicit time integration scheme instead of explicit time integration allows one to use a significantly larger time step without compromising on accuracy, thus yielding a much higher speed of integration. At the same time, using a Fourier spectral method to approximate spatial derivatives gives much higher accuracy than finite difference methods \cite{trefethen}. Thus, the use of a semi-implicit Fourier spectral method enables us to integrate Eq. (\ref{eq: 5}) for long times and large system sizes. The general idea behind the semi-implicit Fourier method is as follows. Consider the following partial differential equation,
\begin{equation} \label{eq: 12}
\frac{\partial u(x,t)}{\partial t}=\hat Lu(x,t) + \hat N[u(x,t), \partial_x u(x,t), \partial_x^2 u(x,t), \dots]
\end{equation}
where $\hat L$ is a linear differential operator and $\hat N$ is a nonlinear functional of $u$ and its spatial derivatives. If we discretize space into $N$ (not to be confused with $\hat N$) grid points with lattice spacing $h$ and denote the value of the field $u(x,t)$ at each grid point by $u_i(t)$, we obtain a system of ordinary differential equations,
\begin{equation} \label{eq: 13}
\frac{du_i}{dt}={(\hat Lu)}_i + \hat N[u(x,t), \dots]_i \ \ \ \ \ \ \ i = 0, 1, \dots, N-1.
\end{equation}
Taking the discrete Fourier transform of this equation gives,
\begin{equation} \label{eq: 14}
\frac{d \widetilde{u}_k}{dt}=\widetilde{(\hat Lu)}_k + \widetilde{\hat N}_k \ \ \ \ \ k=0,1,\dots,\frac{N}{2}-1,-N/2,\dots,-1
\end{equation}
For the (deterministic) SKS equation, 
\begin{equation} \label{eq: 15}
\hat Lu(x,t) = -\alpha u(x,t) -\partial_x^2u(x,t) -\partial_x^4u(x,t)
\end{equation}
and
\begin{equation} \label{eq: 16}
\hat N[u(x,t), \partial_xu(x,t)] = {(\partial_xu(x,t))}^2
\end{equation}
The Fourier transform of the discretized field $u_i$ can be found numerically using the Fast Fourier Transform (FFT). If $\widetilde{u}_k$ denotes the $k$-th wavenumber component of the discrete Fourier transform of the field $u$, then the discrete Fourier transform of Eq. (\ref{eq: 15}) is \cite{trefethen},
\begin{equation} \label{eq: 17}
\widetilde{(\hat Lu)}_k = -\alpha \widetilde{u}_k + {(2\pi k/Nh)}^2 \widetilde{u}_k - {(2\pi k/Nh)}^4 \widetilde{u}_k 
\end{equation}
 Evaluating the nonlinear term $\widetilde{\hat N_k}[u(x,t), \dots]$ in Fourier space directly is computationally expensive, since the Fourier transform of a product of functions is a convolution involving $O(N^2)$ computations. Therefore, we do the integration by evaluating the nonlinear term in position space and then transforming it to Fourier space. Putting all this together, we get,
\begin{equation} \label{eq: 18}
\frac{d \widetilde{u}_k}{dt}= -\alpha \widetilde{u}_k + {(2\pi k/Nh)}^2 \widetilde{u}_k - {(2\pi k/Nh)}^4 \widetilde{u}_k + \widetilde{\hat N}_k[u(x,t), \dots]
\end{equation}
To solve the issue of small time step, we use a semi-implicit integration scheme. We integrate forward in time by treating the linear terms implicitly and the nonlinear term explicitly \cite{chen}. Let $u_k^j$ be the value of $u_k$ at time $t_j$. Then we approximate the time derivative by $du_k^j/dt=\frac{u_k^{j+1}-u_k^j}{\Delta t}$ and Eq. (\ref{eq: 18}) becomes,
\begin{equation} \label{eq: 19}
u_k^{j+1}=\frac{u_k^j + \Delta t  \widetilde{N}_k^j}{1-\Delta t \left[-\alpha +{(\frac{2\pi k}{L})}^2 -{(\frac{2 \pi k}{L})}^4 \right]}
\end{equation}
With this semi-implicit scheme, it is possible to increase the time step by a factor of about 50 compared to that in an explicit time integration scheme, without causing instabilities and without loss of accuracy. This results in a significant speed up of the algorithm. Finally, in order to incorporate noise, we let $\zeta_k^j$ denote the value of the noise term at time $t_j$ and Eq. (\ref{eq: 19}) is modified to read,
\begin{equation} \label{eq: 20}
u_k^{j+1}=\frac{u_k^j + \Delta t  \widetilde{N}_k^j + \sqrt{\frac{2 \varepsilon N \Delta t}{h}} \zeta_k^j}{1-\Delta t \left[-\alpha +{(\frac{2\pi k}{L})}^2 -{(\frac{2 \pi k}{L})}^4 \right]}
\end{equation}

where $\langle \zeta_k^j \rangle=0$ and $\langle \zeta_k^j \zeta_{k^\prime}^{j^\prime} \rangle = \delta_{k,-k^\prime} \delta_{jj^\prime}$.
This is simply the Ito formula for the solution of a stochastic differential equation \cite{garcia}. The procedure for generating noise in Fourier space which satisfies equations Eqs. (\ref{eq: 6}) and (\ref{eq: 7}) is also given in \cite{garcia}, Appendix B.

\begin{table*}[t!]
\caption{\label{table: 2} Most visited wavenumbers for various sizes, $\alpha=0.20$.}
\begin{ruledtabular}
\begin{tabular}{lll}
 
 Number of lattice points $N$ & System Length $(L=Nh)$ & Most visited state \\ 
 \hline
 1024 & 512 & 0.6627 $(n=54)$ \\

 1600 & 800 &   0.6597 $(n=84)$\\ 

 2200 & 1100 & 0.6569 $(n = 115)$\\ 

 3000 & 1500 & 0.6576 $(n = 157)$\\

 4000 & 2000 & 0.6566 $(n = 209)$\\

\end{tabular}
\end{ruledtabular}
\end{table*}

\begin{table*}
\caption{\label{table: 3} Most visited wavenumbers for various sizes, $\alpha=0.17$.}
\begin{ruledtabular}
\begin{tabular}{lll} 

 Number of lattice points $N$ & System Length $(L=Nh)$ & Most visited state \\ 
 \hline
 1024 & 512 & 0.6381 $(n=52)$ \\

 1600 & 800 &  0.6361 $(n=81)$\\
 
 2200 & 1100 &  0.6340 $(n=111)$ \\

 3000 & 1500 & 0.6367 $(n=152)$\\

 4000 & 2000 & 0.6377 $(n=203)$\\
\end{tabular}
\end{ruledtabular}
\end{table*}

\section{Results} \label{res}
Our aim is to compute the empirical probability distribution for the allowed periodic states and determine if the distribution has a peak at a particular wavenumber. If such a peak is present, it would support the hypothesis that noise is a possible mechanism of wavenumber selection. To do this, we use noise strengths that are around 0.1 \% of the amplitude of the field $u(x,t)$. This noise strength is quite large, but it allows the system to explore a large number of allowed states. We start the simulation with the system initially placed in one of the periodic states and then perturb it with noise. The noise causes the system to visit other steady states. We then count the number of times each periodic state is visited and compute the fraction of the time spent in each state, for long runs of $\sim 10^8$ time steps. The idea is that the system will spend the greatest fraction of time in the most stable or selected state. It is important to clarify what we mean by ``visiting a given periodic state''. Since we are using relatively large noise, the state of the system at any given time has a broad power spectrum, with non-zero components for several wavenumbers and it is, strictly speaking, inaccurate to say that the system is in a given periodic state. Instead, we consider the system to be in the neighborhood of a periodic state, if the power spectrum has a peak at that periodicity, and if the Fourier component at this periodicity is at least twice as large as the other Fourier components. Hence, what we actually calculate is the empirical probability for being near a periodic state. To do so, we calculate the time average of the \textit{indicator function} for a state with wavenumber $q$,   
\begin{equation} \label{eq: 21}
 M_T(q) = \frac{1}{T} \int_0^T \mathbf{1}_q(t) \ \text{dt}
 \end{equation}
If the system is in a state with a power spectrum peaked at wavenumber $k$ (and this peak is at least twice as large as other peaks), then the indicator function for wavenumber $q$ is defined by,
 \begin{equation} \label{eq: 22}
 \mathbf{1}_q(t) = 
 \begin{cases}
 1 &  k = q  \ \text{at time} \ t \\
 0 & \text{otherwise}
 \end{cases}
 \end{equation}
 This quantity gives the fraction of time spent near the state with wavenumber $q$ and approaches the stationary probability distribution at very long times.
 \begin{equation} \label{eq: 23}
 \lim_{T \rightarrow \infty} M_T(q) = P_{st}(q)
 \end{equation}
 
Since wavenumber selection has already been demonstrated for a small system with control parameter $\alpha=0.24$ in \cite{obeid}, we attempted to carry out the procedure described above for $\alpha=0.22$. We restrict our simulations to the interval $0.17 \leq \alpha \leq 0.24$, since for $\alpha < 0.16$ the second harmonic in the stationary solution becomes active, leading to complicated instabilities \cite{misbah}. For $\alpha=0.22$, the wave numbers that are stable to perturbations lie within the range $0.6136 \leq q \leq 0.7486$, \cite{misbah}. We used periodic boundary conditions, which implies that out of the wave numbers in the above range, only those appear in the simulation that satisfy,
\begin{equation} \label{eq: 24}
q_n = \frac{2\pi n}{Nh}
\end{equation}
where $n$ is an integer $(1 \leq n \leq N)$. Thus, the imposition of periodic boundary conditions limits the allowed wavenumbers to a discrete set given by Eq. (\ref{eq: 24}), with $n$ representing the number of cells in the solution. For all our simulations, the initial condition was of the form $u_{in} \sim \text{sin}(2\pi n_{in} x/Nh)$ with $n_{in}$ being an integer.
For the range $0.6136 \leq q \leq 0.7486$, and $N=1024$, $h=0.5$, we have $50 \leq n \leq 61$. Our time step was $\Delta t=0.3$. 

\begin{figure*}
 \begin{subfigure}{\linewidth}
   \includegraphics[width=.5\linewidth]{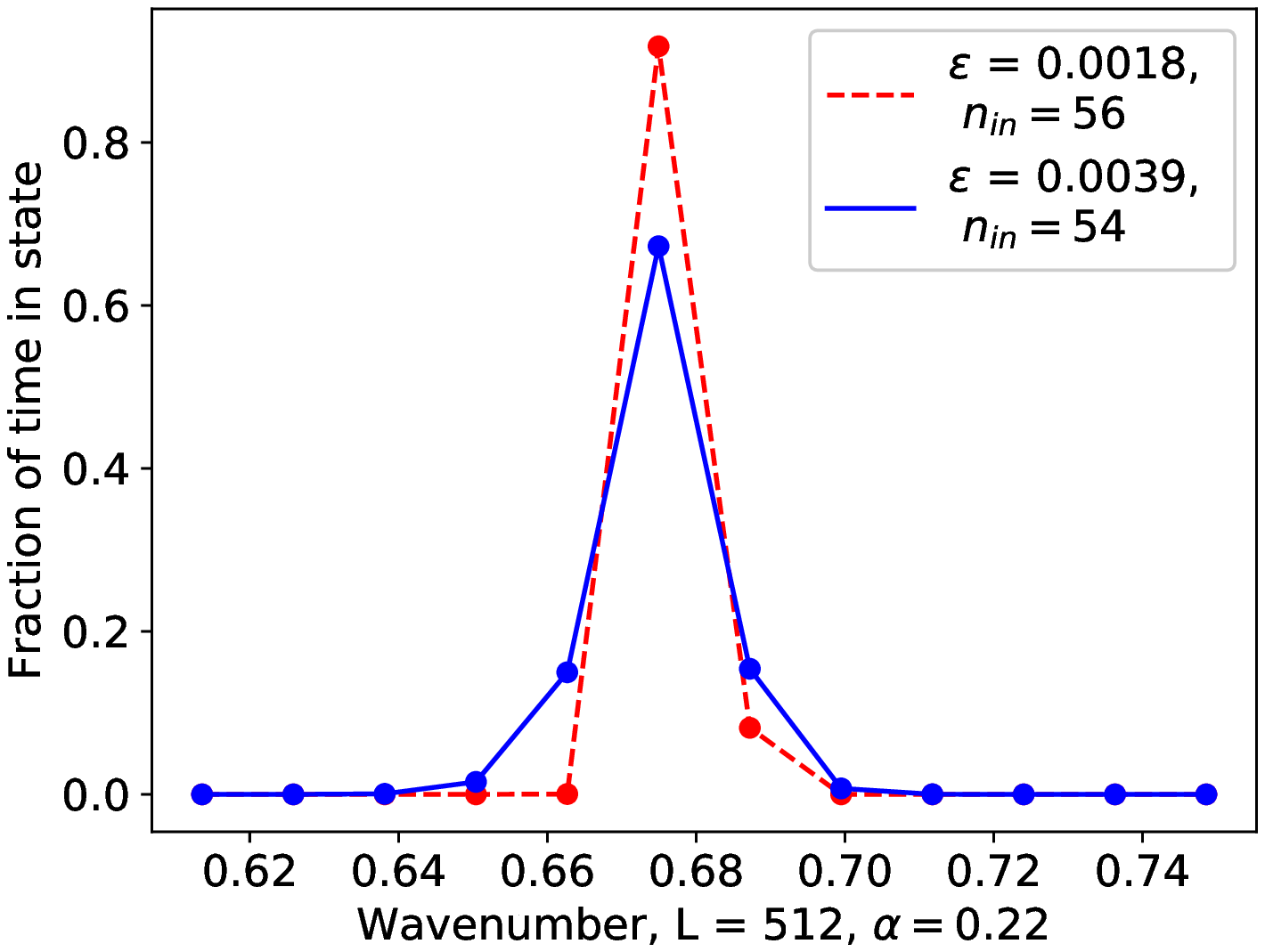}\hfill
  \includegraphics[width=.5\linewidth]{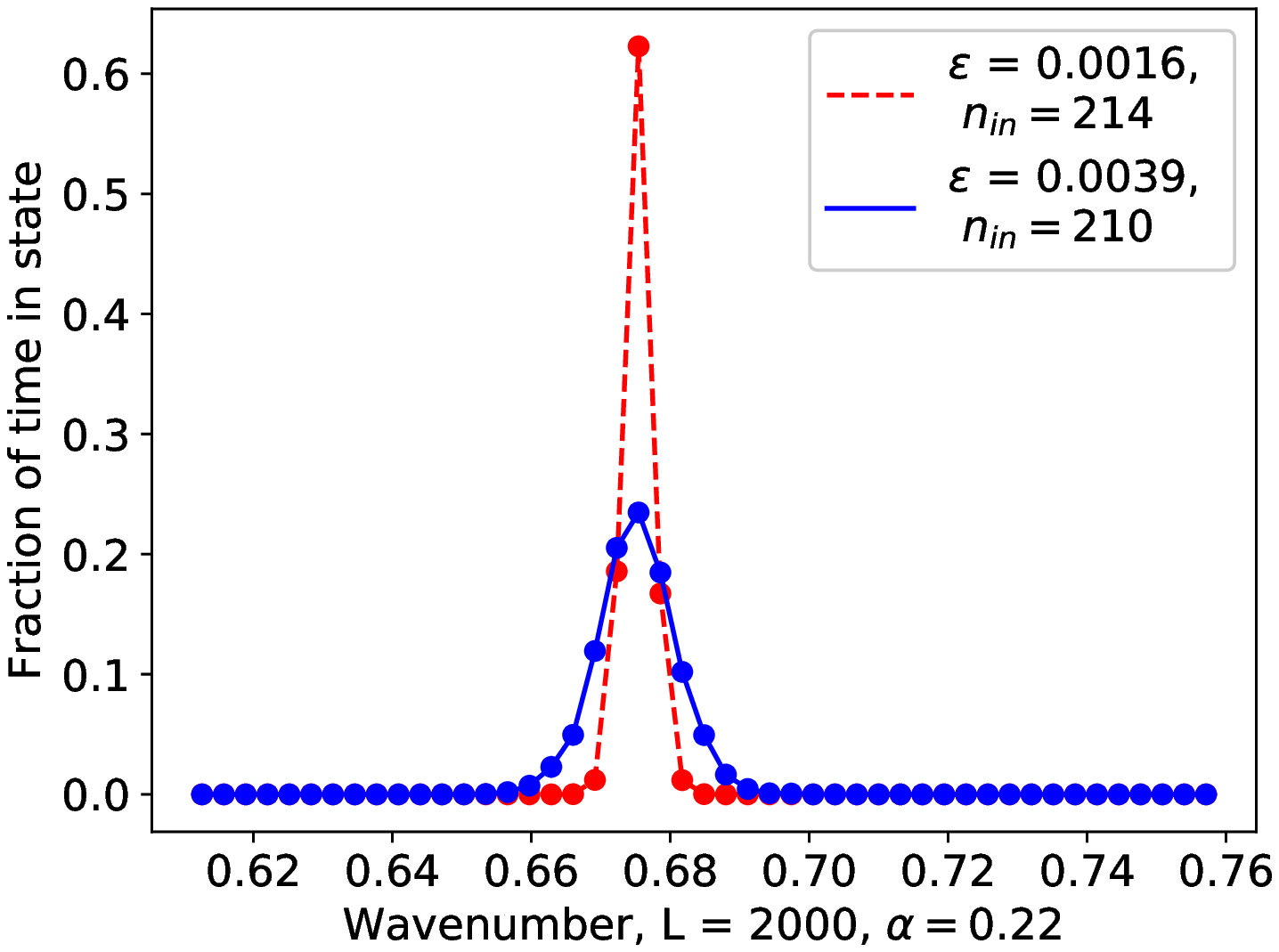}
  \caption{Empirical probability distribution for $\alpha=0.22$ and various noise strengths and initial states, $n_{in}$. Left: $N=1024$. Right: $N=4000$.}
  \end{subfigure} \par \medskip
   \begin{subfigure}{\linewidth}
  \includegraphics[width=.5\linewidth]{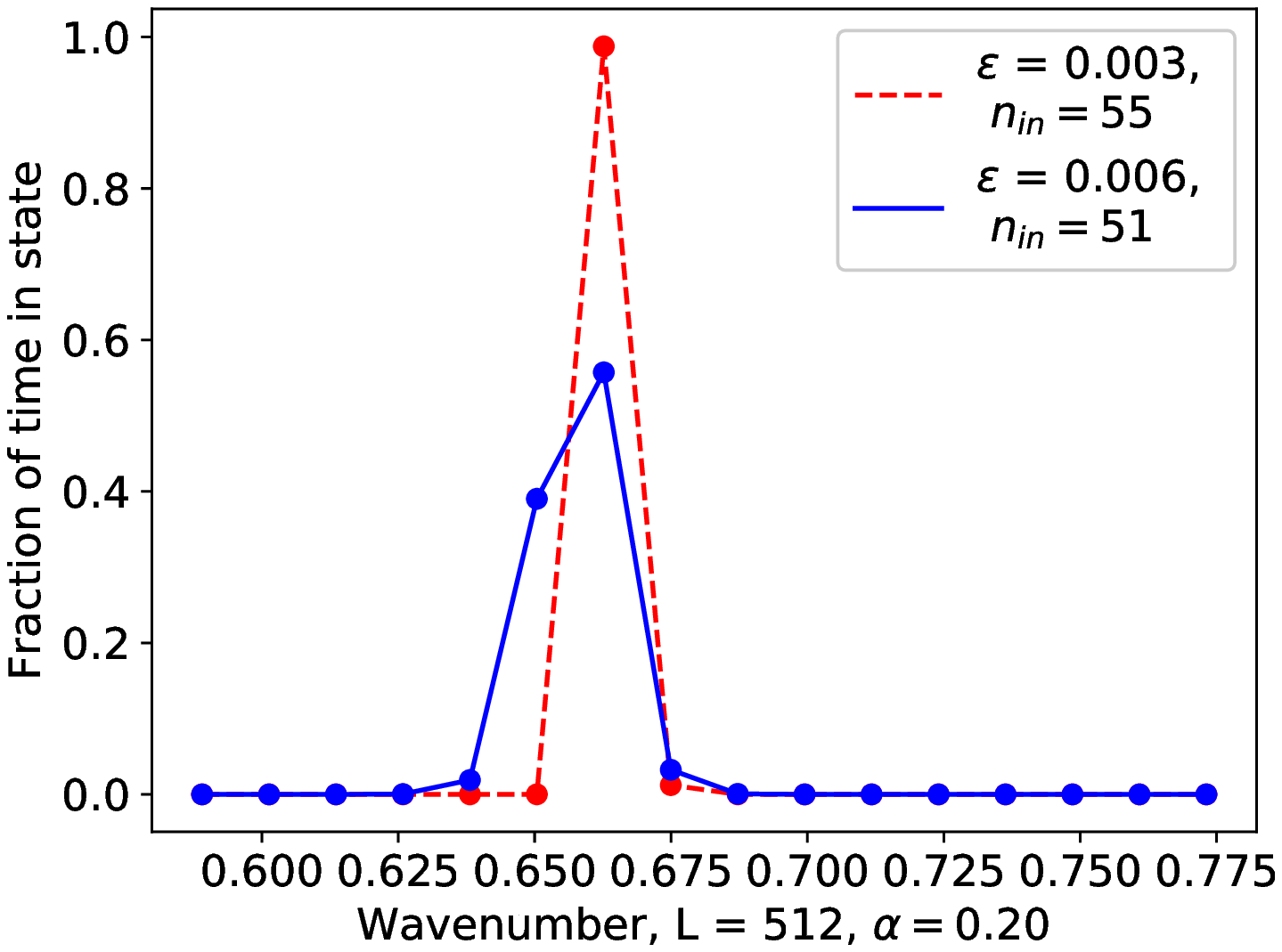}\hfill
  \includegraphics[width=.5\linewidth]{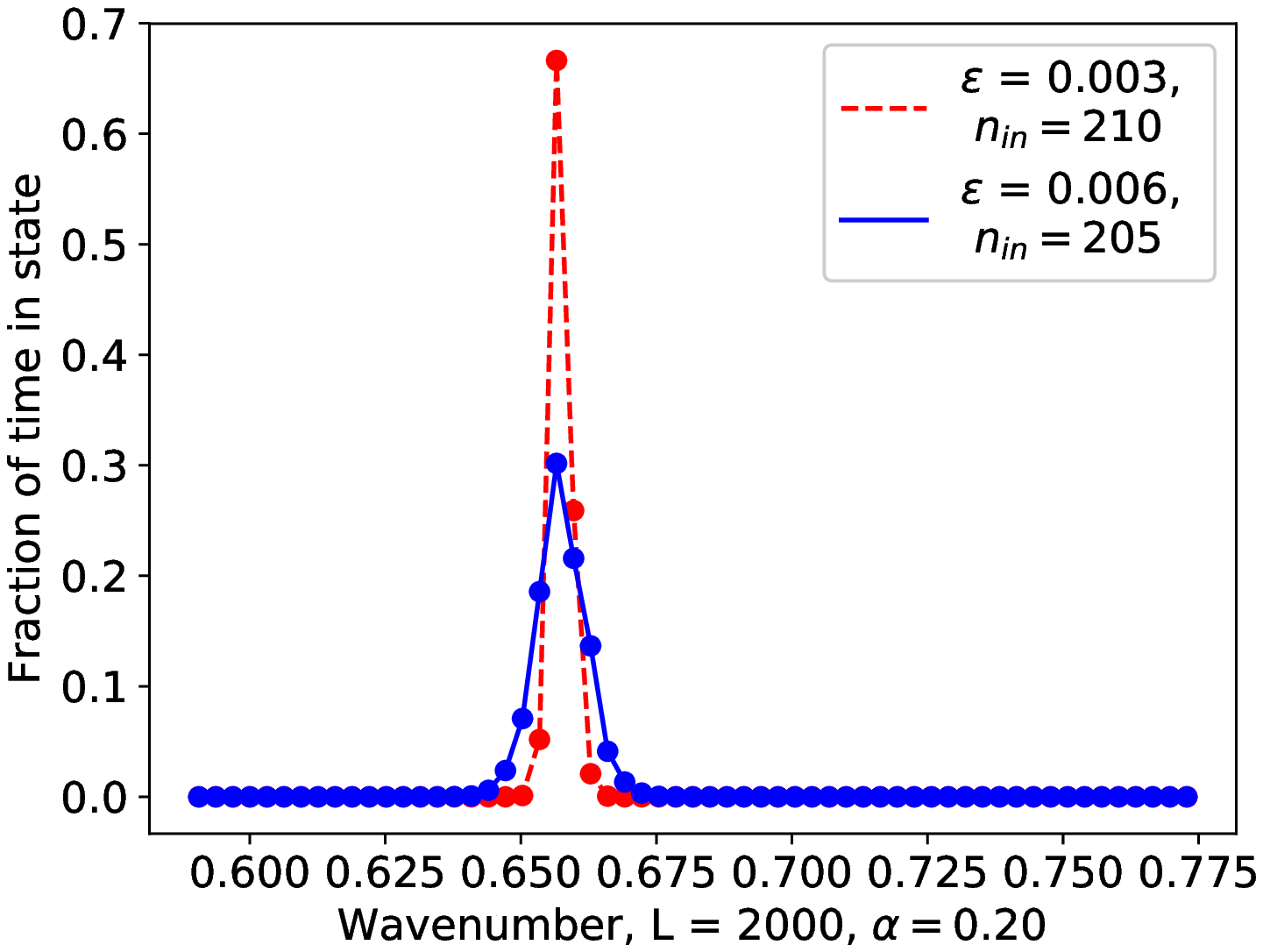}
   \caption{Empirical probability distribution for $\alpha=0.20$ and various noise strengths and initial states, $n_{in}$. Left: $N=1024$. Right: $N=4000$.}
  \end{subfigure} \par \medskip
  \begin{subfigure}{\linewidth}
  \includegraphics[width=.5\linewidth]{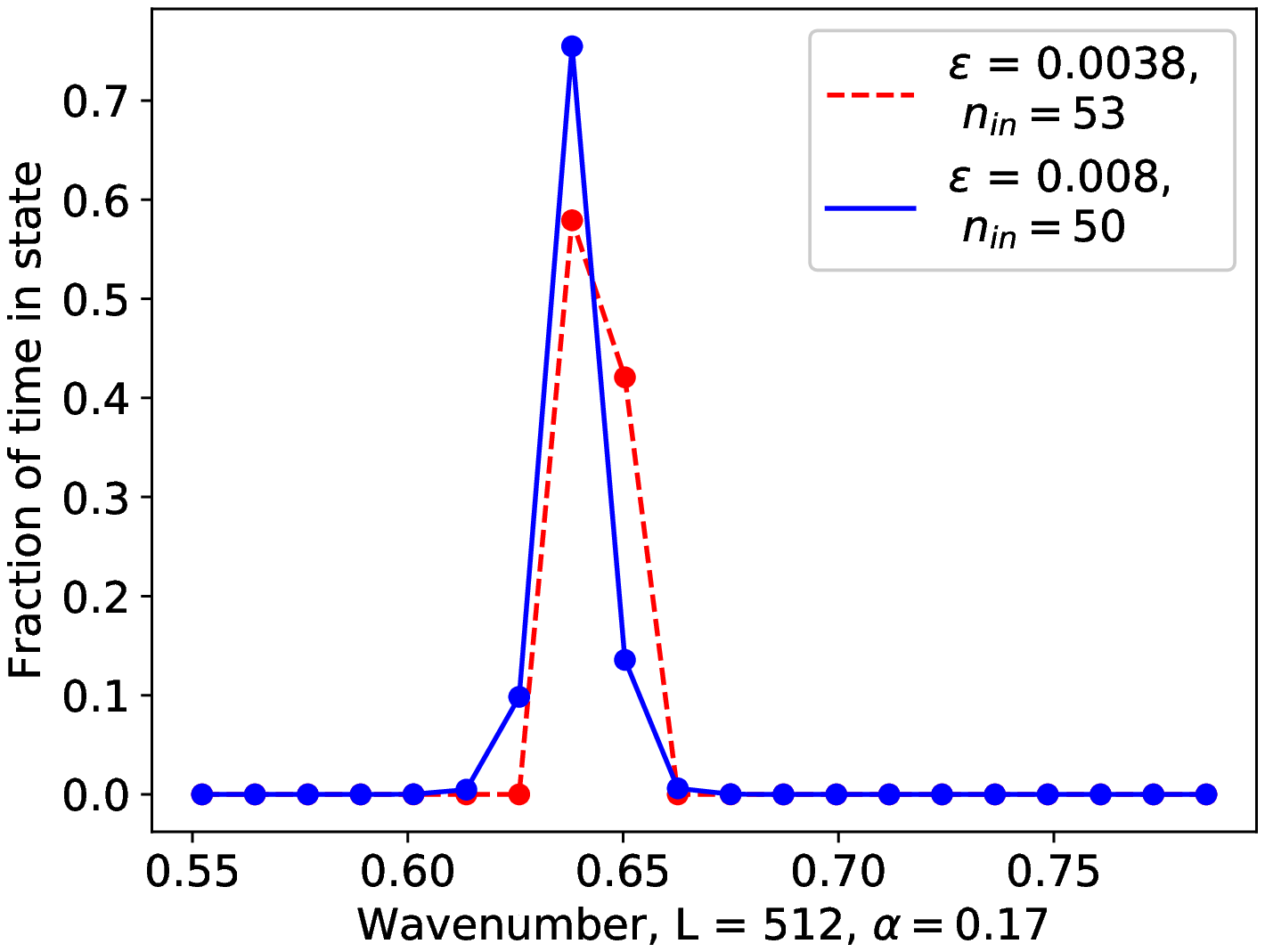}\hfill
  \includegraphics[width=.5\linewidth]{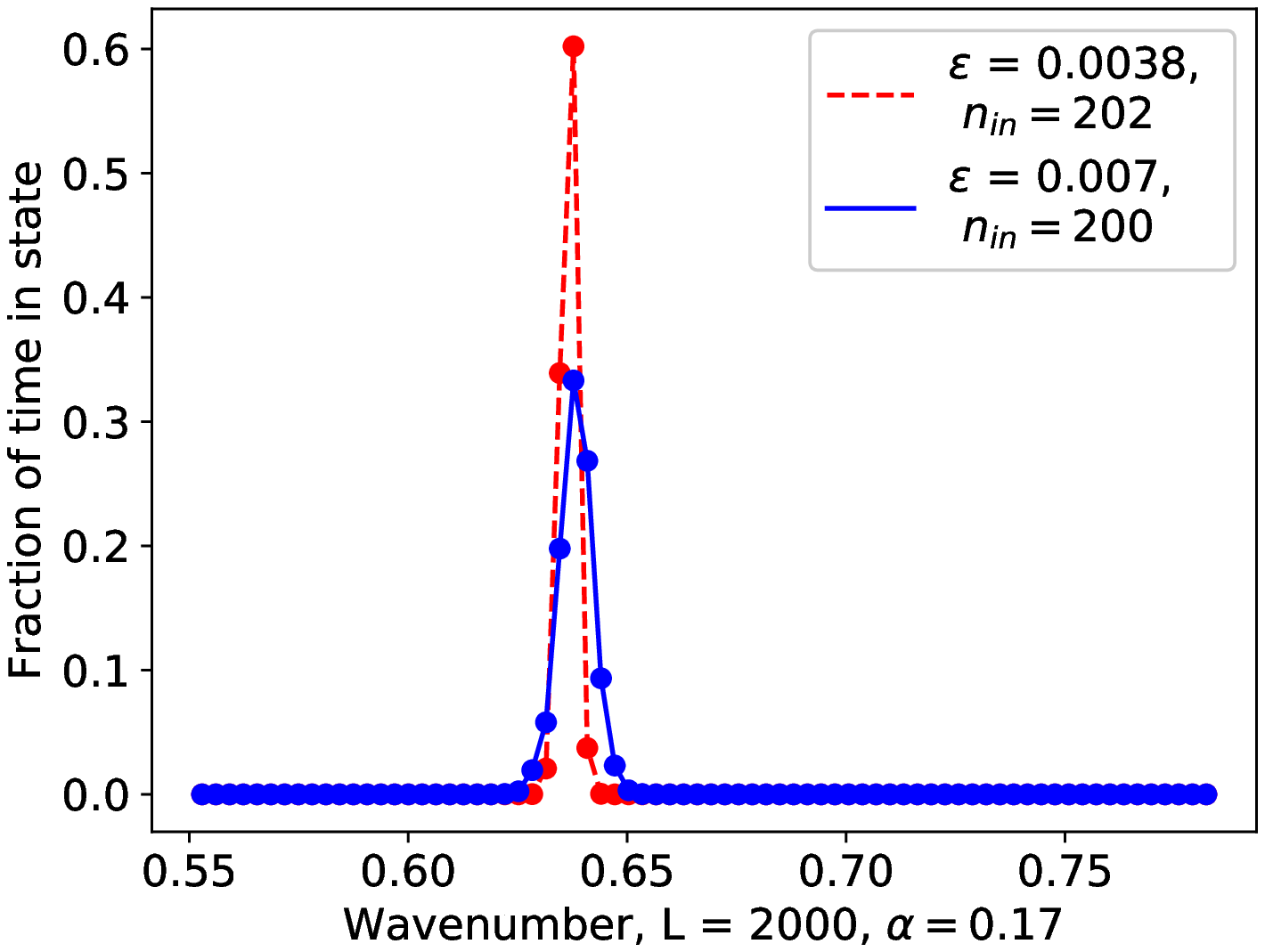}
  \caption{Empirical probability distribution for $\alpha=0.17$ and various noise strengths and initial states, $n_{in}$. Left: $N=1024$. Right: $N=4000$.}
  \end{subfigure}
 \caption{Empirical probability distributions for various control parameter values and small and large system sizes.}
  \label{fig: bigfig1}
\end{figure*}

We first used a small noise strength $\varepsilon=10^{-4}$ to determine which states were the most stable. We found that the states with $n = 54, 55, 56$ were all very stable and the system could not be knocked out of them with this noise strength. So now the problem was to determine which one of these three states was the most stable. We turned up the noise strength gradually and found that at $\varepsilon = 1.8 \times 10^{-3}$, the $n = 54$ and $n=56$ states became unstable, while the $n=55 \ (q=0.6749)$ state persisted. When we plotted the fraction of time spent near each state, the resulting histogram developed a peak at this state. This peak persists until the end of the integration (about $10^8$ time steps), indicating that the system spends most of its time in that state. The histogram also gets narrower with time, again indicating that the system spends most of its time close to the $n=55$ state. We ran a couple of simulations at $\varepsilon=1.9 \times 10^{-3}$ and $\varepsilon=3.9\times 10^{-3}$ and again observed that after spending some time around $n=54$ or $n=56$, the system transitioned to a neighborhood of the $n=55$ state and spent the most amount of time there (see Fig. \ref{fig: bigfig1} (a), left). In all three cases, the histogram became nearly stationary after around $10^8$ time steps.
The observation that the most visited state remains the same in spite of increasing the noise strength is promising evidence of the selection of a unique state by noise. The most visited state was also found to be independent of initial condition. 

We repeated these simulations for the same value of $\alpha$, i.e. $\alpha=0.22$, but different system sizes and noise strengths. In each case, we observed that after initially visiting a narrow band of states, the system settled down close to one of them. This state was independent of the noise strength and initial conditions. The largest system size we tried was $N=4000$ lattice points, which gives us the most precise estimate of the selected wavenumber (see Fig. \ref{fig: bigfig1}(a), right). For this size, the difference between two successive wavenumbers is the smallest and is equal to $\Delta q=2\pi/Nh=0.003$. The most visited states for $\alpha=0.22$ are presented in Table. \ref{table: Table 1}.  Thus, we see that our best estimate for the selected wavenumber for $\alpha=0.22$ is $q_s=0.6754 \pm \Delta q/2$ or $q_s=0.6754 \pm 0.0015$, corresponding to the case $N=4000$. 

Next, we repeated the same process for values of $\alpha$ farther from the critical value $\alpha_c=0.25$. As explained in \cite{obeid}, states in the middle of the Eckhaus band become more and more stable away from threshold. This made it necessary to use slightly larger noise strengths to destabilize the states and obtain the required probability distributions. For $\alpha=0.20$ and $N=1024$, the Eckhaus stable wavenumbers are $0.5890 \leq q \leq 0.7608$ or $48 \leq n \leq 62$. For a range of noise strengths between $3 \times 10^{-3}$ and $6 \times 10^{-3}$ and different initial conditions, we found $n=54 \ (q = 0.6627)$ to be the most visited state (Fig. \ref{fig: bigfig1}(b), left). For $N=4000$ lattice points, we found $n=209$ to be the most visited state, see Fig. \ref{fig: bigfig1}(b), right. Results for various system sizes are shown in Table \ref{table: 2}. Finally, results for $\alpha=0.17$ are shown in Table \ref{table: 3} and Fig. \ref{fig: bigfig1}(c). 

Figs. \ref{fig: bigfig1}(a), (b) and (c) show that the width of the probability distributions shrinks as the system size increases, indicating that the probability distributions approach Dirac delta functions peaked at a unique wavenumber in the thermodynamic ($N \rightarrow \infty$) limit. The probability distributions for each size also become narrower as the noise strength is decreased, but the maximum of the distributions remains at the same wavenumber.
\begin{figure*}
  \begin{subfigure}{\linewidth}
  \includegraphics[width=.5\linewidth]{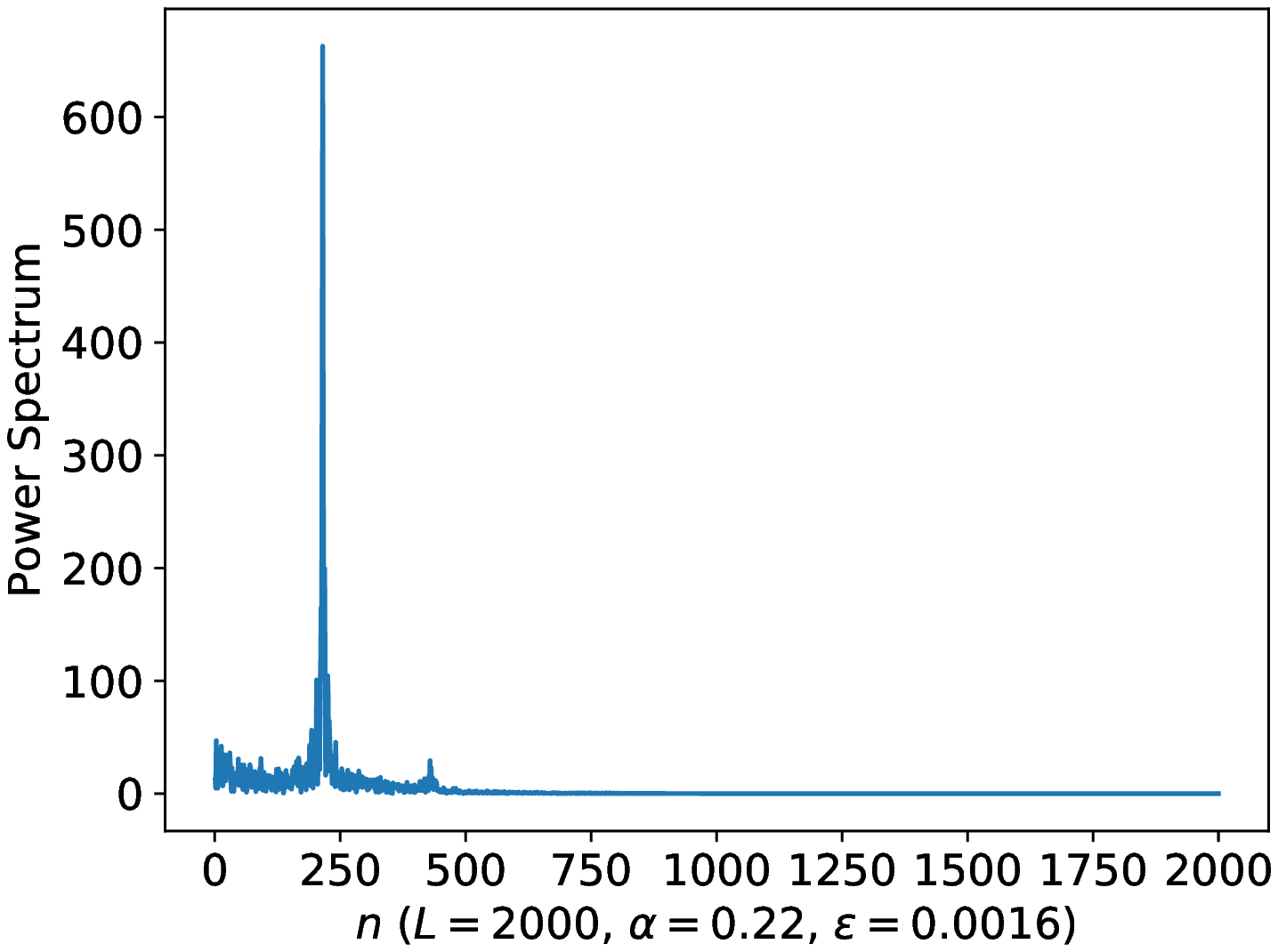}\hfill
   \includegraphics[width=.5\linewidth]{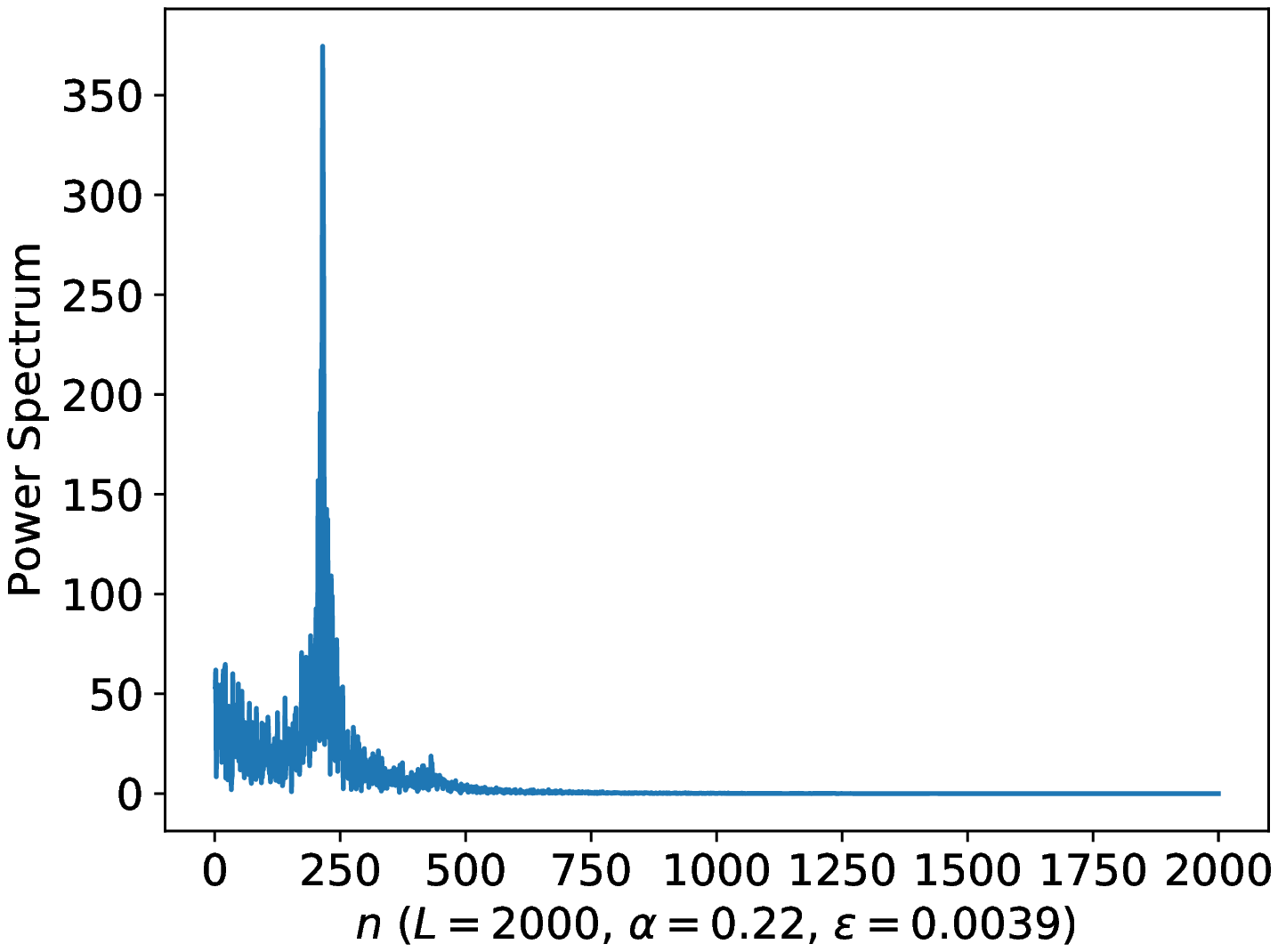}
  \end{subfigure}
    \caption{Long time power spectra for $\alpha=0.22$, 4000 lattice points. Left: $\varepsilon=0.0016$. Right: $\varepsilon=0.0039$.}
    \label{fig: bigfig2}
\end{figure*}
For the sake of completeness, we have also reproduced Obeid's result for $\alpha=0.24$ and extended it to larger sizes, as shown in Table \ref{table: 4}.
\begin{table*}
\caption{\label{table: 4} Most visited wavenumbers for various sizes, $\alpha=0.24$.}
\begin{ruledtabular}
\begin{tabular}{lll} 

 Number of lattice points $N$ & System Length $(L=Nh)$ & Most visited state \\ 
 \hline
 1024 & 512 & 0.6995 $(n=57)$ \\

 1600 & 800 &  0.6990 $(n=89)$\\
 
 2200 & 1100 &  0.6969 $(n=122)$ \\

 3000 & 1500 & 0.6953 $(n=166)$\\

 4000 & 2000 & 0.6974 $(n=222)$\\

\end{tabular}
\end{ruledtabular}
\end{table*}

The power spectra at the end of integration for $\alpha=0.22$ and low and high noise strengths are shown in Fig. \ref{fig: bigfig2}. These figures show that the final power spectra have prominent maxima at the most visited wavenumber, in spite of the large noise strengths we have used. Small but visible second harmonic peaks are also observed, indicating that the system is close to a periodic steady state of the deterministic system. The power spectra for $\alpha=0.20$ and $0.17$ are similar. Note that we have plotted the number of cells $n$ on the x-axis, instead of the wavenumber.
\begin{figure*}
  \begin{subfigure}{\linewidth}
  \includegraphics[width=.5\linewidth]{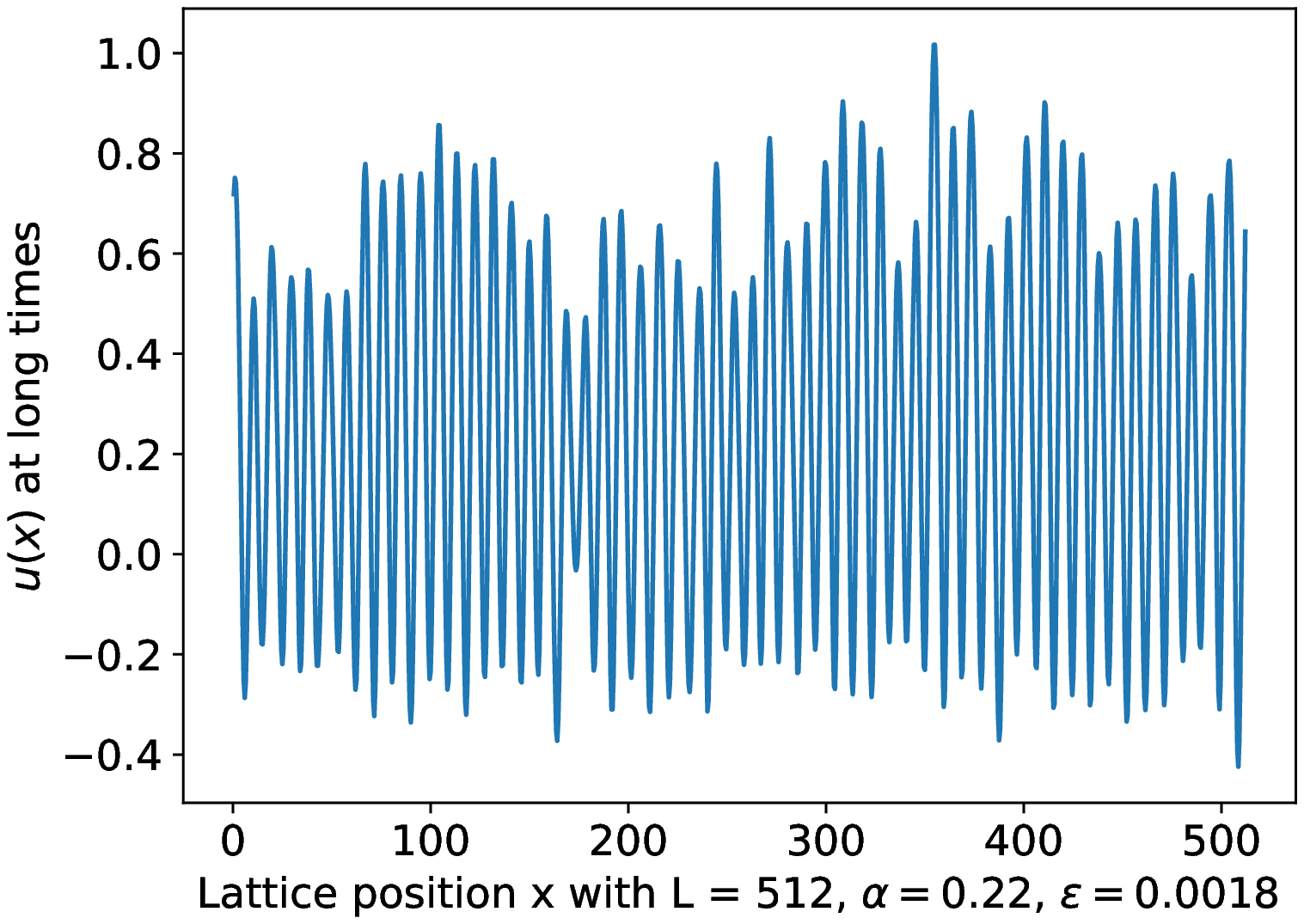}\hfill
  \includegraphics[width=.5\linewidth]{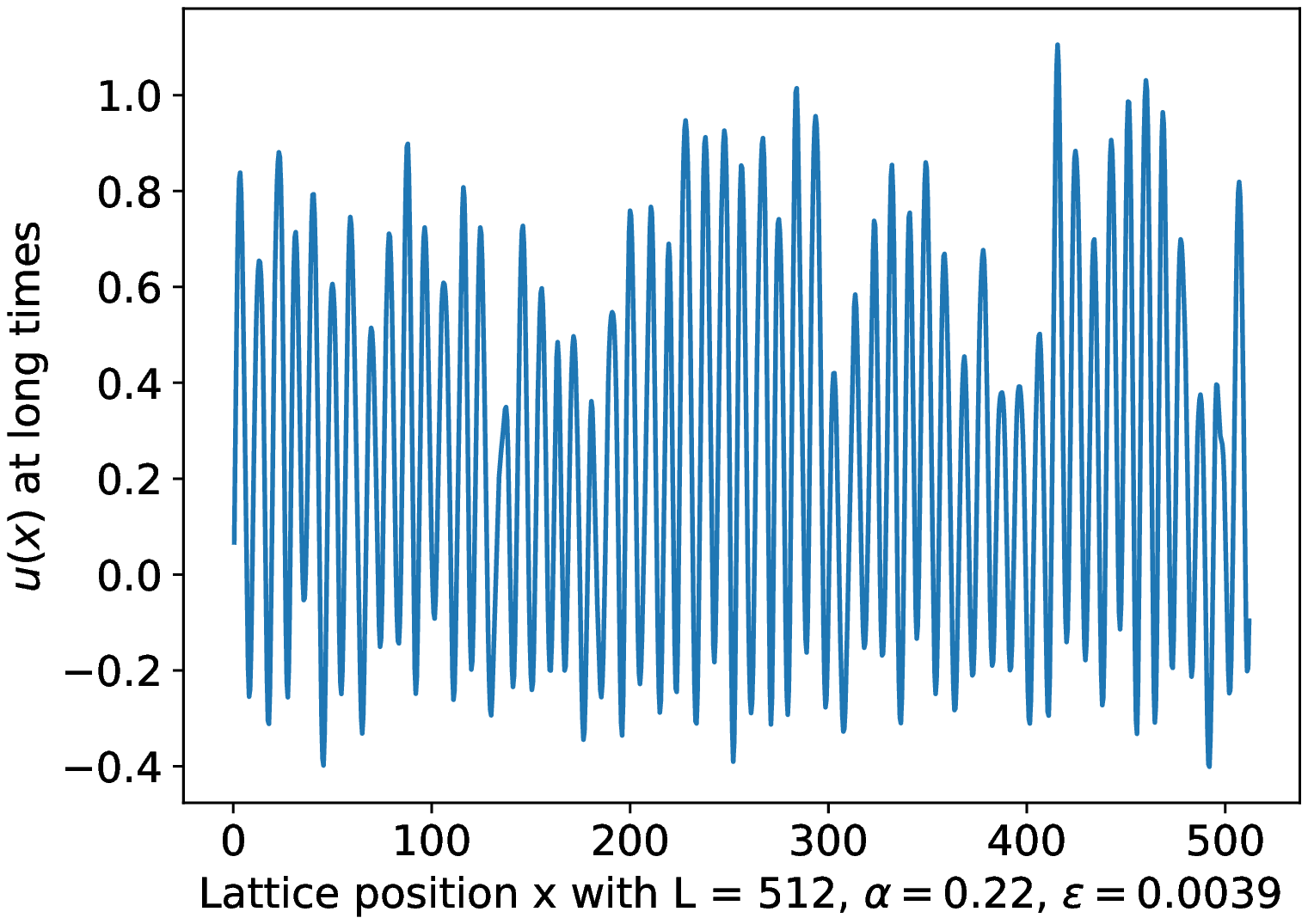}
 \end{subfigure}
  \caption{Typical final configurations at the end of integration for $\alpha=0.22, 1024$ lattice points, for low and high noise strengths.}
  \label{fig: bigfig3}
\end{figure*}

\begin{figure}
\centering
\includegraphics[width=0.48\textwidth]{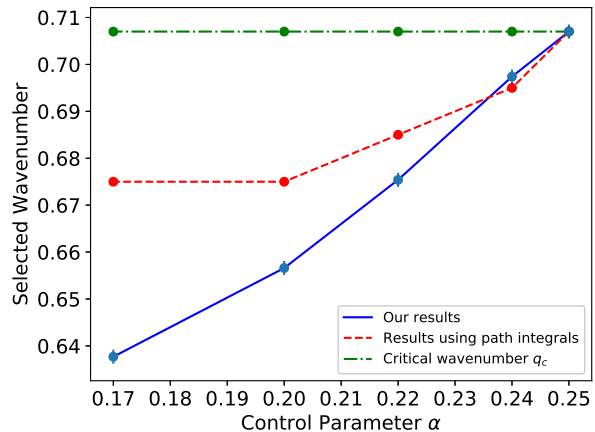}
\caption{Comparison of our results with those of \cite{qiao}. The horizontal line represents the critical wavenumber $q_c=\frac{1}{\sqrt{2}}$, which is the fastest growing wavenumber in the linear stability analysis.}
\label{fig: comparison}
\end{figure}

We also show plots of the field $u(x,t)$ (for $\alpha=0.22$ and 1024 lattice points) at very long times in Fig. \ref{fig: bigfig3}. The left plot is for $\varepsilon=0.0018$, while the right plot is for $\varepsilon=0.0039$. Finally, a plot of the selected wavenumbers for 4000 lattice points against $\alpha$ is shown in Fig. \ref{fig: comparison}, with error bars representing the discretization error. It also shows the results obtained in \cite{qiao} for the same values of $\alpha$. Both curves show that as the value of $\alpha$ is decreased below threshold, the selected wavenumber is shifted to the left of the critical wavenumber $q_c$ (horizontal line in Fig. \ref{fig: comparison}). However, our values for the selected wavenumber do not agree with theirs far from $\alpha=\alpha_c$. We believe that the reason for the disagreement is likely numerical. Since the disagreement is particularly large away from threshold, it is our view that the use of the amplitude equation may be inappropriate. It is possible, for example, that the saddle solution for the amplitude equation is not a sufficiently accurate initial guess for the true saddle state for $\alpha$ values far from 0.25. Secondly, the time reversed deterministic path is the minimum action path only for potential systems \cite{qiao}. While it is reasonable to assume that it would be a good initial guess for non-potential systems, it is also possible that the actual minimum action path is far more complicated away from threshold, where the non-potential term in Eq. (\ref{eq: 5}) is not small. The discretized action is a function of the discretized field at all lattice points and all times, and hence could be an extremely complicated function, with possibly multiple minima. It is not clear what kind of numerical uncertainties arise in minimizing this action, and whether the minimum action paths found in \cite{qiao} are local minima or global minima. A thorough analysis of the optimization techniques is needed to explore these issues.

 \begin{figure}
\centering
\includegraphics[width=0.48\textwidth]{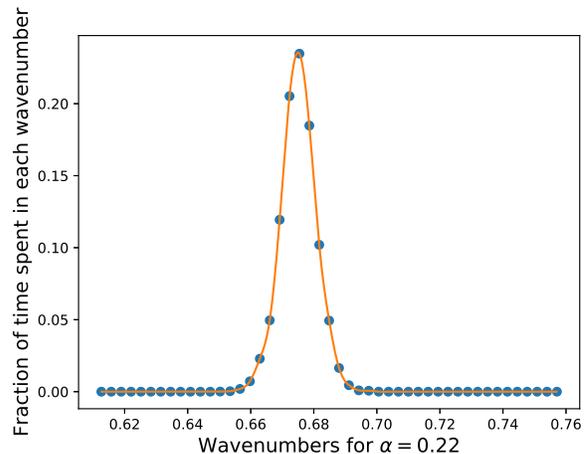}
\caption{Interpolating curve, $\alpha=0.22$, with a maximum at $q_s=0.6748$.}
\label{fig: interp022}
\end{figure}

 \begin{figure*}
    \centering
    \begin{minipage}{0.48\textwidth}
        \centering
        \includegraphics[width=\textwidth]{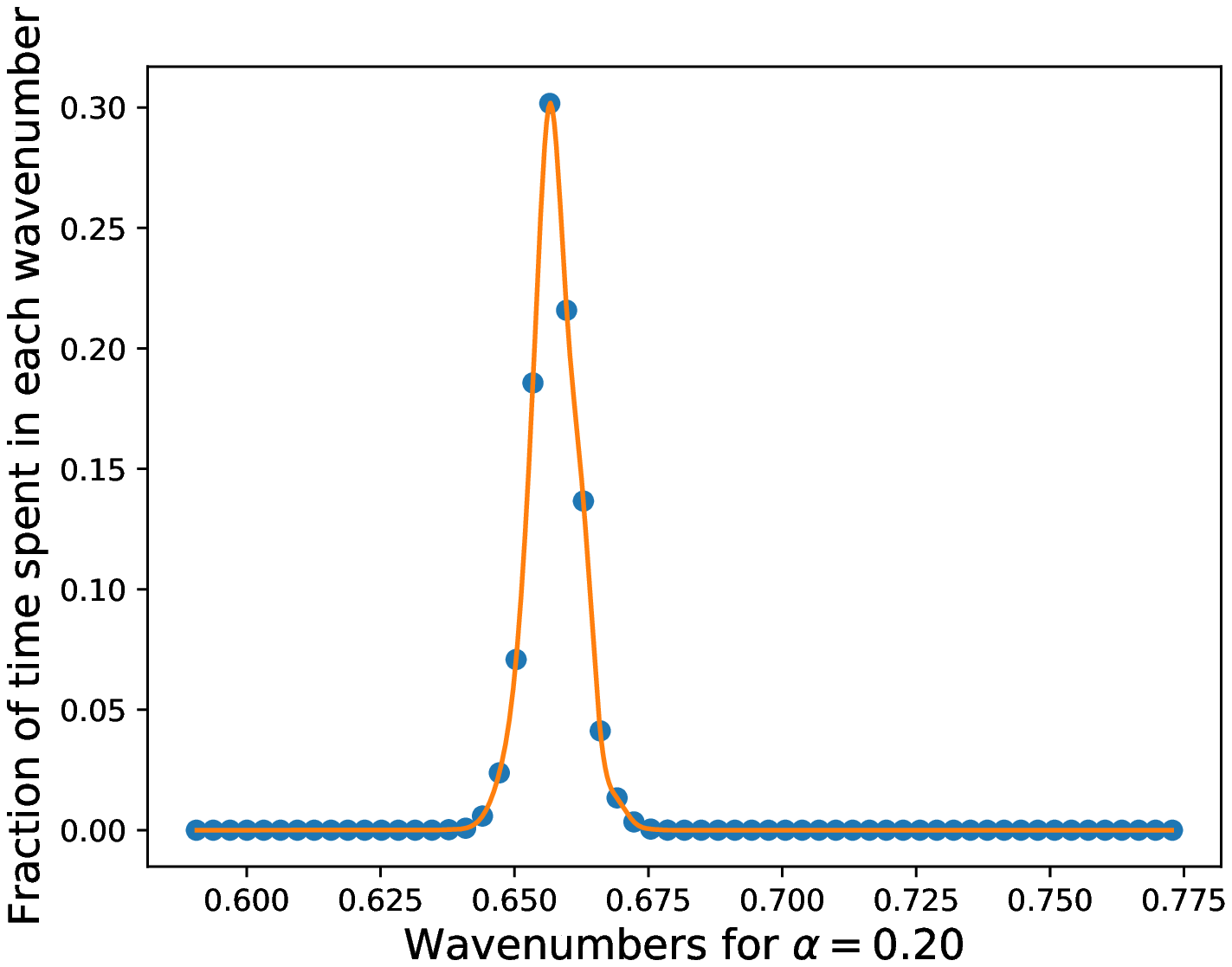} 
        \caption{Interpolating curve, $\alpha=0.20$, with a maximum at $q_s=0.6567$.}
        \label{figure: interp020}
    \end{minipage}
    \begin{minipage}{0.48\textwidth}
        \centering
        \includegraphics[width=\textwidth]{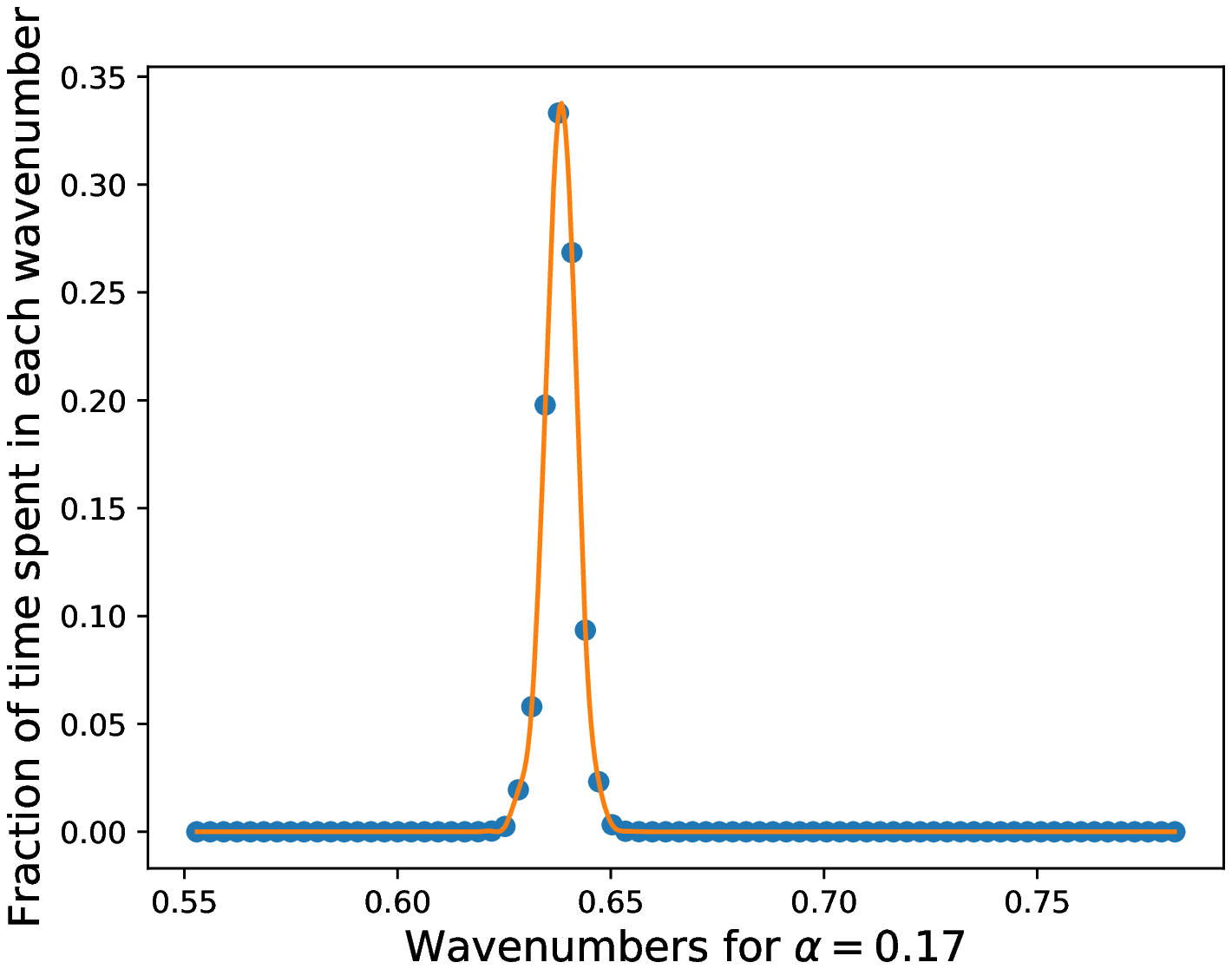} 
        \caption{Interpolating curve, $\alpha=0.17$ with a maximum at $q_s=0.6384$.}
        \label{fig: interp017}
    \end{minipage}
\end{figure*}
 
 \subsection{Extension to thermodynamic limit}
  In order to obtain an estimate for the selected wavenumber for the thermodynamic limit (in which case a continuous band of wavenumbers is allowed), we used cubic spline interpolation to get a smooth function passing through the discrete points obtained from our simulations. We implemented this procedure for $\alpha=0.22, 0.20$ and $0.17$ and $N=4000$ lattice points. The continuous curves thus obtained show which wavenumber would be most probable in the continuum case. The results are shown in Figs. \ref{fig: interp022} through \ref{fig: interp017}. Our continuum estimates for the selected wavenumbers are given in Table \ref{table: 5}.

\begin{table*}
\caption{\label{table: 5} Most visited or selected wavenumbers $q_s$ in the thermodynamic limit.}
\begin{ruledtabular}
\begin{tabular}{lll} 

Control Parameter $\alpha$ & $q_s$  \\ 
 \hline
 0.22 & 0.6748\\

 0.20 & 0.6567\\
 
 0.17 & 0.6384\\
\end{tabular}
\end{ruledtabular}
\end{table*}

\section{Conclusions} \label{conc}
We have performed a detailed numerical study of noise induced wavenumber selection in the one dimensional SKS equation. We have shown that even in the presence of large noise, long time power spectra of the field are most likely to be peaked at a unique wavenumber that does not depend on the initial state or the noise strength. We note again that due to the large noise strengths we had to use, there were non-zero Fourier components at other wavenumbers, which means that for finite system sizes and large noise strengths, there is no true selection, as expected. However, since the position of the peak of the power spectrum is independent of noise strength and since the probability distributions become sharper with increasing system size and decreasing noise strength, it is plausible to conclude that the same wavenumber would be selected for noise strengths much less than the ones used here, except that it would require prohibitively large integration times.

It is interesting to compare our findings with other selection mechanisms proposed in \cite{cann, kramer, riecke}. As discussed there, control parameter ramps are an efficient way to select a unique, well defined wavenumber. However, \cite{riecke} also shows that varying different quantities that lead to the same final control parameter results in different selected wavenumbers. Ref. \cite{schober} has shown that starting from random initial conditions and simulating deterministically, the final wavenumber depends on the initial state. In contrast, our work shows that at long times, the system is most likely to be found in a state with a dominant wavenumber, and this dominant wavenumber does not depend on initial conditions or noise strength, but is an intrinsic property of the SKS system. We have been able to show this for a wide range of control parameters and large system sizes, in contrast to previous work. Our work shows that noise induced wavenumber selection can occur in non-potential systems. Much remains to be done however; a future direction could be to devise efficient importance sampling techniques which would enable us to study transitions between states without using large noise strengths. It would also be interesting to see if it is possible to predict the selected wavenumber analytically, although to our knowledge no such analytical theory exists.


\begin{thebibliography}{23}%
\makeatletter
\providecommand \@ifxundefined [1]{%
 \@ifx{#1\undefined}
}%
\providecommand \@ifnum [1]{%
 \ifnum #1\expandafter \@firstoftwo
 \else \expandafter \@secondoftwo
 \fi
}%
\providecommand \@ifx [1]{%
 \ifx #1\expandafter \@firstoftwo
 \else \expandafter \@secondoftwo
 \fi
}%
\providecommand \natexlab [1]{#1}%
\providecommand \enquote  [1]{``#1''}%
\providecommand \bibnamefont  [1]{#1}%
\providecommand \bibfnamefont [1]{#1}%
\providecommand \citenamefont [1]{#1}%
\providecommand \href@noop [0]{\@secondoftwo}%
\providecommand \href [0]{\begingroup \@sanitize@url \@href}%
\providecommand \@href[1]{\@@startlink{#1}\@@href}%
\providecommand \@@href[1]{\endgroup#1\@@endlink}%
\providecommand \@sanitize@url [0]{\catcode `\\12\catcode `\$12\catcode
  `\&12\catcode `\#12\catcode `\^12\catcode `\_12\catcode `\%12\relax}%
\providecommand \@@startlink[1]{}%
\providecommand \@@endlink[0]{}%
\providecommand \url  [0]{\begingroup\@sanitize@url \@url }%
\providecommand \@url [1]{\endgroup\@href {#1}{\urlprefix }}%
\providecommand \urlprefix  [0]{URL }%
\providecommand \Eprint [0]{\href }%
\providecommand \doibase [0]{http://dx.doi.org/}%
\providecommand \selectlanguage [0]{\@gobble}%
\providecommand \bibinfo  [0]{\@secondoftwo}%
\providecommand \bibfield  [0]{\@secondoftwo}%
\providecommand \translation [1]{[#1]}%
\providecommand \BibitemOpen [0]{}%
\providecommand \bibitemStop [0]{}%
\providecommand \bibitemNoStop [0]{.\EOS\space}%
\providecommand \EOS [0]{\spacefactor3000\relax}%
\providecommand \BibitemShut  [1]{\csname bibitem#1\endcsname}%
\let\auto@bib@innerbib\@empty
\bibitem [{\citenamefont {Cross}\ and\ \citenamefont
  {Hohenberg}(1993)}]{cross}%
  \BibitemOpen
  \bibfield  {author} {\bibinfo {author} {\bibfnamefont {M.~C.}\ \bibnamefont
  {Cross}}\ and\ \bibinfo {author} {\bibfnamefont {P.~C.}\ \bibnamefont
  {Hohenberg}},\ }\href@noop {} {\bibfield  {journal} {\bibinfo  {journal}
  {Rev. Mod. Phys.}\ }\textbf {\bibinfo {volume} {65}},\ \bibinfo {pages} {851}
  (\bibinfo {year} {1993})}\BibitemShut {NoStop}%
\bibitem [{\citenamefont {Grossmann}\ \emph {et~al.}(1993)\citenamefont
  {Grossmann}, \citenamefont {Elder}, \citenamefont {Grant},\ and\
  \citenamefont {Kosterlitz}}]{dirsol}%
  \BibitemOpen
  \bibfield  {author} {\bibinfo {author} {\bibfnamefont {B.}~\bibnamefont
  {Grossmann}}, \bibinfo {author} {\bibfnamefont {K.~R.}\ \bibnamefont
  {Elder}}, \bibinfo {author} {\bibfnamefont {M.}~\bibnamefont {Grant}}, \ and\
  \bibinfo {author} {\bibfnamefont {J.~M.}\ \bibnamefont {Kosterlitz}},\
  }\href@noop {} {\bibfield  {journal} {\bibinfo  {journal} {Phys. Rev. Lett.}\
  }\textbf {\bibinfo {volume} {71}},\ \bibinfo {pages} {3323} (\bibinfo {year}
  {1993})}\BibitemShut {NoStop}%
\bibitem [{\citenamefont {Cross}\ and\ \citenamefont {Greenside}(2009)}]{crgs}%
  \BibitemOpen
  \bibfield  {author} {\bibinfo {author} {\bibfnamefont {M.~C.}\ \bibnamefont
  {Cross}}\ and\ \bibinfo {author} {\bibfnamefont {H.}~\bibnamefont
  {Greenside}},\ }\href@noop {} {\emph {\bibinfo {title} {Pattern Formation and
  Dynamics in Nonequilibrium Systems}}}\ (\bibinfo  {publisher} {Cambridge
  University Press},\ \bibinfo {address} {New York},\ \bibinfo {year}
  {2009})\BibitemShut {NoStop}%
\bibitem [{\citenamefont {Schober}\ \emph {et~al.}(1986)\citenamefont
  {Schober}, \citenamefont {Allroth}, \citenamefont {Schroeder},\ and\
  \citenamefont {M\"uller-Krumbhaar}}]{schober}%
  \BibitemOpen
  \bibfield  {author} {\bibinfo {author} {\bibfnamefont {H.~R.}\ \bibnamefont
  {Schober}}, \bibinfo {author} {\bibfnamefont {E.}~\bibnamefont {Allroth}},
  \bibinfo {author} {\bibfnamefont {K.}~\bibnamefont {Schroeder}}, \ and\
  \bibinfo {author} {\bibfnamefont {H.}~\bibnamefont {M\"uller-Krumbhaar}},\
  }\href@noop {} {\bibfield  {journal} {\bibinfo  {journal} {Phys. Rev. A}\
  }\textbf {\bibinfo {volume} {33}},\ \bibinfo {pages} {567} (\bibinfo {year}
  {1986})}\BibitemShut {NoStop}%
\bibitem [{\citenamefont {Cannell}\ \emph {et~al.}(1983)\citenamefont
  {Cannell}, \citenamefont {Dominguez-Lerma},\ and\ \citenamefont
  {Ahlers}}]{cann}%
  \BibitemOpen
  \bibfield  {author} {\bibinfo {author} {\bibfnamefont {D.~S.}\ \bibnamefont
  {Cannell}}, \bibinfo {author} {\bibfnamefont {M.~A.}\ \bibnamefont
  {Dominguez-Lerma}}, \ and\ \bibinfo {author} {\bibfnamefont {G.}~\bibnamefont
  {Ahlers}},\ }\href@noop {} {\bibfield  {journal} {\bibinfo  {journal} {Phys.
  Rev. Lett.}\ }\textbf {\bibinfo {volume} {50}},\ \bibinfo {pages} {1365}
  (\bibinfo {year} {1983})}\BibitemShut {NoStop}%
\bibitem [{\citenamefont {Kramer}\ \emph {et~al.}(1982)\citenamefont {Kramer},
  \citenamefont {Ben-Jacob}, \citenamefont {Brand},\ and\ \citenamefont
  {Cross}}]{kramer}%
  \BibitemOpen
  \bibfield  {author} {\bibinfo {author} {\bibfnamefont {L.}~\bibnamefont
  {Kramer}}, \bibinfo {author} {\bibfnamefont {E.}~\bibnamefont {Ben-Jacob}},
  \bibinfo {author} {\bibfnamefont {H.}~\bibnamefont {Brand}}, \ and\ \bibinfo
  {author} {\bibfnamefont {M.~C.}\ \bibnamefont {Cross}},\ }\href@noop {}
  {\bibfield  {journal} {\bibinfo  {journal} {Phys. Rev. Lett.}\ }\textbf
  {\bibinfo {volume} {49}},\ \bibinfo {pages} {1891} (\bibinfo {year}
  {1982})}\BibitemShut {NoStop}%
\bibitem [{\citenamefont {Hohenberg}\ \emph {et~al.}(1985)\citenamefont
  {Hohenberg}, \citenamefont {Kramer},\ and\ \citenamefont {Riecke}}]{riecke}%
  \BibitemOpen
  \bibfield  {author} {\bibinfo {author} {\bibfnamefont {P.~C.}\ \bibnamefont
  {Hohenberg}}, \bibinfo {author} {\bibfnamefont {L.}~\bibnamefont {Kramer}}, \
  and\ \bibinfo {author} {\bibfnamefont {H.}~\bibnamefont {Riecke}},\
  }\href@noop {} {\bibfield  {journal} {\bibinfo  {journal} {Physica}\ }\textbf
  {\bibinfo {volume} {15D}},\ \bibinfo {pages} {402} (\bibinfo {year}
  {1985})}\BibitemShut {NoStop}%
\bibitem [{\citenamefont {Kerszberg}(1983{\natexlab{a}})}]{kers1}%
  \BibitemOpen
  \bibfield  {author} {\bibinfo {author} {\bibfnamefont {M.}~\bibnamefont
  {Kerszberg}},\ }\href@noop {} {\bibfield  {journal} {\bibinfo  {journal}
  {Phys. Rev. B}\ }\textbf {\bibinfo {volume} {27}},\ \bibinfo {pages} {3909}
  (\bibinfo {year} {1983}{\natexlab{a}})}\BibitemShut {NoStop}%
\bibitem [{\citenamefont {Kerszberg}(1983{\natexlab{b}})}]{kers2}%
  \BibitemOpen
  \bibfield  {author} {\bibinfo {author} {\bibfnamefont {M.}~\bibnamefont
  {Kerszberg}},\ }\href@noop {} {\bibfield  {journal} {\bibinfo  {journal}
  {Phys. Rev. B}\ }\textbf {\bibinfo {volume} {28}},\ \bibinfo {pages} {247}
  (\bibinfo {year} {1983}{\natexlab{b}})}\BibitemShut {NoStop}%
\bibitem [{\citenamefont {Filho}\ \emph {et~al.}(2005)\citenamefont {Filho},
  \citenamefont {Kosterlitz},\ and\ \citenamefont {Granato}}]{enzo}%
  \BibitemOpen
  \bibfield  {author} {\bibinfo {author} {\bibfnamefont {R.~N.~C.}\
  \bibnamefont {Filho}}, \bibinfo {author} {\bibfnamefont {J.~M.}\ \bibnamefont
  {Kosterlitz}}, \ and\ \bibinfo {author} {\bibfnamefont {E.}~\bibnamefont
  {Granato}},\ }\href@noop {} {\bibfield  {journal} {\bibinfo  {journal}
  {Physica A}\ }\textbf {\bibinfo {volume} {354}},\ \bibinfo {pages} {333}
  (\bibinfo {year} {2005})}\BibitemShut {NoStop}%
\bibitem [{\citenamefont {Obeid}\ \emph {et~al.}(2010)\citenamefont {Obeid},
  \citenamefont {Kosterlitz},\ and\ \citenamefont {Sandstede}}]{obeid}%
  \BibitemOpen
  \bibfield  {author} {\bibinfo {author} {\bibfnamefont {D.}~\bibnamefont
  {Obeid}}, \bibinfo {author} {\bibfnamefont {J.~M.}\ \bibnamefont
  {Kosterlitz}}, \ and\ \bibinfo {author} {\bibfnamefont {B.}~\bibnamefont
  {Sandstede}},\ }\href@noop {} {\bibfield  {journal} {\bibinfo  {journal}
  {Phys. Rev. E}\ }\textbf {\bibinfo {volume} {81}},\ \bibinfo {pages} {066205}
  (\bibinfo {year} {2010})}\BibitemShut {NoStop}%
\bibitem [{\citenamefont {Vi{\~{n}}als}\ \emph {et~al.}(1991)\citenamefont
  {Vi{\~{n}}als}, \citenamefont {Hern\'andez-Garc\'ia}, \citenamefont {san
  Miguel},\ and\ \citenamefont {Toral}}]{vinals}%
  \BibitemOpen
  \bibfield  {author} {\bibinfo {author} {\bibfnamefont {J.}~\bibnamefont
  {Vi{\~{n}}als}}, \bibinfo {author} {\bibfnamefont {E.}~\bibnamefont
  {Hern\'andez-Garc\'ia}}, \bibinfo {author} {\bibfnamefont {M.}~\bibnamefont
  {San Miguel}}, \ and\ \bibinfo {author} {\bibfnamefont {R.}~\bibnamefont
  {Toral}},\ }\href@noop {} {\bibfield  {journal} {\bibinfo  {journal} {Phys.
  Rev. A}\ }\textbf {\bibinfo {volume} {44}},\ \bibinfo {pages} {1123}
  (\bibinfo {year} {1991})}\BibitemShut {NoStop}%
\bibitem [{\citenamefont {Swift}\ and\ \citenamefont
  {Hohenberg}(1977)}]{swift}%
  \BibitemOpen
  \bibfield  {author} {\bibinfo {author} {\bibfnamefont {J.}~\bibnamefont
  {Swift}}\ and\ \bibinfo {author} {\bibfnamefont {P.~C.}\ \bibnamefont
  {Hohenberg}},\ }\href@noop {} {\bibfield  {journal} {\bibinfo  {journal}
  {Phys. Rev. A}\ }\textbf {\bibinfo {volume} {15}},\ \bibinfo {pages} {319}
  (\bibinfo {year} {1977})}\BibitemShut {NoStop}%
\bibitem [{\citenamefont {Misbah}\ and\ \citenamefont
  {Valance}(1994)}]{misbah}%
  \BibitemOpen
  \bibfield  {author} {\bibinfo {author} {\bibfnamefont {C.}~\bibnamefont
  {Misbah}}\ and\ \bibinfo {author} {\bibfnamefont {A.}~\bibnamefont
  {Valance}},\ }\href@noop {} {\bibfield  {journal} {\bibinfo  {journal} {Phys.
  Rev. E}\ }\textbf {\bibinfo {volume} {49}},\ \bibinfo {pages} {166} (\bibinfo
  {year} {1994})}\BibitemShut {NoStop}%
\bibitem [{\citenamefont {Qiao}\ \emph {et~al.}(2016)\citenamefont {Qiao},
  \citenamefont {Zheng},\ and\ \citenamefont {Cross}}]{qiao}%
  \BibitemOpen
  \bibfield  {author} {\bibinfo {author} {\bibfnamefont {L.}~\bibnamefont
  {Qiao}}, \bibinfo {author} {\bibfnamefont {Z.}~\bibnamefont {Zheng}}, \ and\
  \bibinfo {author} {\bibfnamefont {M.~C.}\ \bibnamefont {Cross}},\ }\href@noop
  {} {\bibfield  {journal} {\bibinfo  {journal} {Phys. Rev. E}\ }\textbf
  {\bibinfo {volume} {93}},\ \bibinfo {pages} {042204} (\bibinfo {year}
  {2016})}\BibitemShut {NoStop}%
\bibitem [{\citenamefont {Bena}\ \emph {et~al.}(1993)\citenamefont {Bena},
  \citenamefont {Misbah},\ and\ \citenamefont {Valance}}]{misbah2}%
  \BibitemOpen
  \bibfield  {author} {\bibinfo {author} {\bibfnamefont {I.}~\bibnamefont
  {Bena}}, \bibinfo {author} {\bibfnamefont {C.}~\bibnamefont {Misbah}}, \ and\
  \bibinfo {author} {\bibfnamefont {A.}~\bibnamefont {Valance}},\ }\href@noop
  {} {\bibfield  {journal} {\bibinfo  {journal} {Phys. Rev. B}\ }\textbf
  {\bibinfo {volume} {47}},\ \bibinfo {pages} {7408} (\bibinfo {year}
  {1993})}\BibitemShut {NoStop}%
\bibitem [{\citenamefont {Garc\'ia-Ojalvo}\ \emph {et~al.}(1993)\citenamefont
  {Garc\'ia-Ojalvo}, \citenamefont {Hern\'andez-Machado},\ and\ \citenamefont
  {Sancho}}]{ghs93}%
  \BibitemOpen
  \bibfield  {author} {\bibinfo {author} {\bibfnamefont {J.}~\bibnamefont
  {Garc\'ia-Ojalvo}}, \bibinfo {author} {\bibfnamefont {A.}~\bibnamefont
  {Hern\'andez-Machado}}, \ and\ \bibinfo {author} {\bibfnamefont {J.~M.}\
  \bibnamefont {Sancho}},\ }\href@noop {} {\bibfield  {journal} {\bibinfo
  {journal} {Phys. Rev. Lett.}\ }\textbf {\bibinfo {volume} {71}},\ \bibinfo
  {pages} {1542} (\bibinfo {year} {1993})}\BibitemShut {NoStop}%
\bibitem [{\citenamefont {Parrondo}\ \emph {et~al.}(1996)\citenamefont
  {Parrondo}, \citenamefont {den Broeck}, \citenamefont {Buceta},\ and\
  \citenamefont {de~la Rubia}}]{pvbr96}%
  \BibitemOpen
  \bibfield  {author} {\bibinfo {author} {\bibfnamefont {J.~M.~R.}\
  \bibnamefont {Parrondo}}, \bibinfo {author} {\bibfnamefont {C.~V.}\
  \bibnamefont {den Broeck}}, \bibinfo {author} {\bibfnamefont
  {J.}~\bibnamefont {Buceta}}, \ and\ \bibinfo {author} {\bibfnamefont
  {J.}~\bibnamefont {de~la Rubia}},\ }\href@noop {} {\bibfield  {journal}
  {\bibinfo  {journal} {Physica A}\ }\textbf {\bibinfo {volume} {224}},\
  \bibinfo {pages} {153} (\bibinfo {year} {1996})}\BibitemShut {NoStop}%
\bibitem [{\citenamefont {Freidlin}\ and\ \citenamefont
  {Wentzell}(2012)}]{freidlin}%
  \BibitemOpen
  \bibfield  {author} {\bibinfo {author} {\bibfnamefont {M.~I.}\ \bibnamefont
  {Freidlin}}\ and\ \bibinfo {author} {\bibfnamefont {A.~D.}\ \bibnamefont
  {Wentzell}},\ }\href@noop {} {\emph {\bibinfo {title} {Random Perturbations
  of Dynamical Systems}}}\ (\bibinfo  {publisher} {Springer},\ \bibinfo
  {address} {New York},\ \bibinfo {year} {2012})\BibitemShut {NoStop}%
\bibitem [{\citenamefont {Kramer}\ and\ \citenamefont
  {Zimmermann}(1985)}]{zimm}%
  \BibitemOpen
  \bibfield  {author} {\bibinfo {author} {\bibfnamefont {L.}~\bibnamefont
  {Kramer}}\ and\ \bibinfo {author} {\bibfnamefont {W.}~\bibnamefont
  {Zimmermann}},\ }\href@noop {} {\bibfield  {journal} {\bibinfo  {journal}
  {Physica D}\ }\textbf {\bibinfo {volume} {16}},\ \bibinfo {pages} {221}
  (\bibinfo {year} {1985})}\BibitemShut {NoStop}%
\bibitem [{\citenamefont {Chen}\ and\ \citenamefont {Shen}(1998)}]{chen}%
  \BibitemOpen
  \bibfield  {author} {\bibinfo {author} {\bibfnamefont {L.~Q.}\ \bibnamefont
  {Chen}}\ and\ \bibinfo {author} {\bibfnamefont {J.}~\bibnamefont {Shen}},\
  }\href@noop {} {\bibfield  {journal} {\bibinfo  {journal} {Comput. Phys.
  Commun.}\ }\textbf {\bibinfo {volume} {108}},\ \bibinfo {pages} {147}
  (\bibinfo {year} {1998})}\BibitemShut {NoStop}%
\bibitem [{\citenamefont {Trefethen}(2000)}]{trefethen}%
  \BibitemOpen
  \bibfield  {author} {\bibinfo {author} {\bibfnamefont {L.~N.}\ \bibnamefont
  {Trefethen}},\ }\href@noop {} {\emph {\bibinfo {title} {Spectral Methods in
  MATLAB}}}\ (\bibinfo  {publisher} {SIAM},\ \bibinfo {address}
  {Philadelphia},\ \bibinfo {year} {2000})\BibitemShut {NoStop}%
\bibitem [{\citenamefont {Garcia-Ojalvo}\ and\ \citenamefont
  {Sancho}(1999)}]{garcia}%
  \BibitemOpen
  \bibfield  {author} {\bibinfo {author} {\bibfnamefont {J.}~\bibnamefont
  {Garcia-Ojalvo}}\ and\ \bibinfo {author} {\bibfnamefont {J.~M.}\ \bibnamefont
  {Sancho}},\ }\href@noop {} {\emph {\bibinfo {title} {Noise in Spatially
  Extended Systems}}}\ (\bibinfo  {publisher} {Springer-Verlag},\ \bibinfo
  {address} {New York},\ \bibinfo {year} {1999})\BibitemShut {NoStop}%
\end{thebibliography}

%

\end{document}